\documentclass[format=sigconf, preprint]{acmart}
\pdfoutput=1

\usepackage{gensymb}

\usepackage{float}
\usepackage{subcaption}
\usepackage{color}
\usepackage{amsmath}
\usepackage{url}
\usepackage{algorithm}
\usepackage{adjustbox}
\usepackage{enumitem}
\usepackage{wrapfig}
\usepackage[justification=centering]{caption}
\usepackage[noend]{algpseudocode}
\usepackage{setspace}
\usepackage{caption}
\usepackage{multirow}
\newcommand{\DAP}[1]{device activation patterns}
\usepackage{tabto}

\NumTabs{6}
\makeatletter
\def\BState{\State\hskip-\ALG@thistlm}
\makeatother

\newcommand{\fbdr}{flexibility-based DR }

\usepackage{multicol}

\renewcommand\footnotetextcopyrightpermission[1]{} 
\begin{document}
\title{Utilizing Device-level Demand Forecasting for Flexibility Markets - Full Version}

\author{Bijay Neupane}
\affiliation{%
  \institution{Department of Computer Science}
  \city{Aalborg University, Denmark} 
}
\email{bn21@cs.aau.dk}

\author{Torben Bach Pedersen}
\affiliation{%
  \institution{Department of Computer Science}
  \city{Aalborg University, Denmark}
}
\email{tbp@cs.aau.dk}

\author{Bo Thiesson}
\affiliation{%
  \institution{Enversion A/S}
  \city{Aarhus, Denmark}
}
\email{thiesson@enversion.dk}

\begin{abstract}
The uncertainty in the power supply due to fluctuating Renewable Energy Sources (RES) has severe (financial and other) implications for energy market players. In this paper, we present a device-level Demand Response (DR) scheme that captures the atomic (all available) flexibilities in energy demand and provides the largest possible solution space to generate demand/supply schedules that minimize market imbalances. We evaluate the effectiveness and feasibility of widely used forecasting models for device-level flexibility analysis.  In a typical device-level flexibility forecast, a market player is more concerned with the \textit{utility} that the demand flexibility brings to the market, rather than the intrinsic forecast accuracy. In this regard, we provide comprehensive predictive modeling and scheduling of demand flexibility from household appliances to demonstrate the (financial and otherwise) viability of introducing flexibility-based DR in the Danish/Nordic market. Further, we investigate the correlation between the potential utility and the accuracy of the demand forecast model. Furthermore, we perform a number of experiments to determine the data granularity that provides the best financial reward to market players for adopting the proposed DR scheme. A cost-benefit analysis of forecast results shows that even with somewhat low forecast accuracy, market players can achieve regulation cost savings of 54\% of the theoretically optimal.
\end{abstract}
\maketitle

\footnotetext{To appear in the Proceedings of the e-Energy 2018, ninth ACM International Conference on Future Energy Systems (ACM e-Energy 2018) }
\section{Introduction}
The increasing integration of fluctuating Renewable Energy Sources (RES) has an immense impact on the energy market balance in the Nordic region, which has traditionally been maintained by trading a significant volume of energy in the regulation market or by shutting down some of the RES. In either case, the market players lose a substantial amount of revenue in the effort to balance the deviation in expected supply/demand. 
Consequently, there have been various smart grid projects aiming at the efficient utilization of intermittent RES production, and the markets have adopted various Demand Response (DR) programs \cite{Siano2014}, e.g., price-based DR \cite{Zhuang2013, Rad2010, Roscoe2010}, demand reduction bidding \cite{Albadi2007, Aalami2010}, load shift strategy \cite{Biegel2013, Sundstrom2012, Papadaskalopoulos2013}, etc. Further, techniques for integrating household devices into demand-side management for leveling of fluctuating RES production has been explored in \cite{Luns2010, Bigler2011, Li2017, Setlhaolo2014, Adika2014}. Currently, there has been a lot of attention towards flexibility-based DR technique with a focus on utilizing the demand flexibility to control electricity consumption actively, e.g., \cite{Totalflex, Gottwalt2017, Sajjad2016, GEORGES2017, MOHSENI2017}.

In particular, the concept of directly controlling and capturing the shiftable portion of energy demand/supply from individual (IoT-enabled) devices to generate a dynamic schedule that minimizes market imbalance is promising \cite{Totalflex, mirabel, Boehm2012}. Here, the atomic (device-level) flexibilities are explicit and provide the largest possible solution space to generate effective demand and supply schedules. In comparison, higher-level flexibilities, e.g., at the household or feeder level, are too ambiguous to allow optimal aggregation and scheduling due to missing information about the actual source of flexibility.  The mandatory requirement for Distribution System Operators (DSOs) to install smart meters in all Danish households and the introduction of smart devices has enabled an avenue of realizing flexibility-based DR.  Market players can utilize flexibility to compensate their deviation from a baseline or delaying huge investments in grid capacity.  At the same time, consumers can participate in the flexibility-based DR contributing flexibilities in their device usage in response to financial or other incentives.

Most of the proposed flexibility-based DR schemes rely on explicit user input on flexibility information which  requires frequent user involvement and thus lead to response fatigue (reduced participation) in the long run \cite{Kim2011}. 
Hence, for effective implementation of flexibility-based DR, accurate and timely predictions of both non-shiftable and shiftable energy demands are vital.  Further, if we consider energy management for households, a prognosis of device-level demand is fundamental for optimal scheduling of devices to reduce the CO2 emission and lower the energy bill \cite{Barbato2011}. Moreover, the concept of utilizing demand flexibility to obtain a dynamic energy balance and the latency required to support the implementation of the concept (for flexibility extraction and scheduling) make the projection of future device-level demands indispensable.

However, the stochasticity associated with device-level demand makes utilization of the traditional forecast models for device-level demand forecasting a challenging task. Nevertheless, a market player is always more interested in the utility that device-level demand flexibility brings to the market (the financial value of the flexibility) rather than the intrinsic model-level quality (accuracy) of the forecast model. Although there have been some works on quantification of the benefit of load shifting \cite{Pruggler2013}, the analyses are based on markets with less integration of RES compared to Denmark where RES fulfills more than 40\% of the electricity demand. Further, most of the previous work focuses on the quantification of the reduction in the customer's energy bill \cite{Vasirani2013}. However, in a grid system with higher percentages of RES, it is rather players like Balance Responsible Party (BRP) and Distribution System Operator(DSO) that generate substantial savings by avoiding the regulation market and network congestions, respectively.

With the overall goal of assessing device level forecast model, evaluating the financial viability of flexibility-based DR, and quantifying the effect of forecast errors, this paper makes following contributions:\\
\textbf{1.} We assess the accuracy and feasibility of widely used forecasting model, i.e., Logistic Regression for device-level demand forecasting. We present a number of device-level features crafted to capture device usage patterns reliably. Further, we investigate the data granularity and forecast model best suited for the device-level forecasting and flexibility-based DR. Furthermore, we present a rule of thumb for setting the model parameters value that give a near-optimal solution.\\
\textbf{2.} We formulate a set of equations for quantifying the financial benefits of flexibility in energy demand and the loss due to forecast errors. The overall benefit and loss are decomposed and analyzed based on types of prediction categories, i.e. true positive, false positive, etc. Thereupon, we evaluate the best configuration of device-level demand forecast that maximizes the benefit of flexibility-based DR.\\
\textbf{3.} We show that the performance of a classification model improves with data aggregation, and the model achieves the best Area under the precision-recall curve of 0.85 and 0.23 for the daily and hourly resolution, respectively. We further show that even with the lower accuracy for hourly resolution, a market can achieve up to 54\% of the theoretically optimal savings in regulation cost.

The rest of the paper is organized as follows. Section 2 describes the concepts of demand flexibility and flex-offers. Section 3 provides information on the Nordic regulating power market. Section 4 presents our demand forecast technique. Section 5 presents the econometric assessment of flexibility-based DR in relation to the forecast error. Section 6 presents the experimental results and analysis. Finally, Sections 7 concludes the paper and provides directions for future research.

\section{Demand Flexibility}
Demand flexibility refers to the possibility of preponing or postponing some or all of electricity demands from consumption (and production) devices, satisfying user imposed and other constraints. Flexibility is represented in two dimensions, (i) \textit{time flexibility}, the time range within which the demand can be shifted and ii) \textit{amount flexibility}, the range between maximum and minimum demand. For example, if we consider a demand at the device-level, say, a \textit{washer dryer}, \textit{time flexibility} represents the possibility of shifting the activation time to better match an anticipated surplus production from RES. Similarly, the \textit{amount flexibility} represents the volume of energy demand from, say, electric heating that can be scaled up or down according to the market requirement. The potential of extracting flexibilities from the usage of household devices is demonstrated in \cite{Bijay2014}.

\begin{figure}
	\centering
	\includegraphics[width=.7\columnwidth, , trim=0px 0px 0px 0px, clip=true]{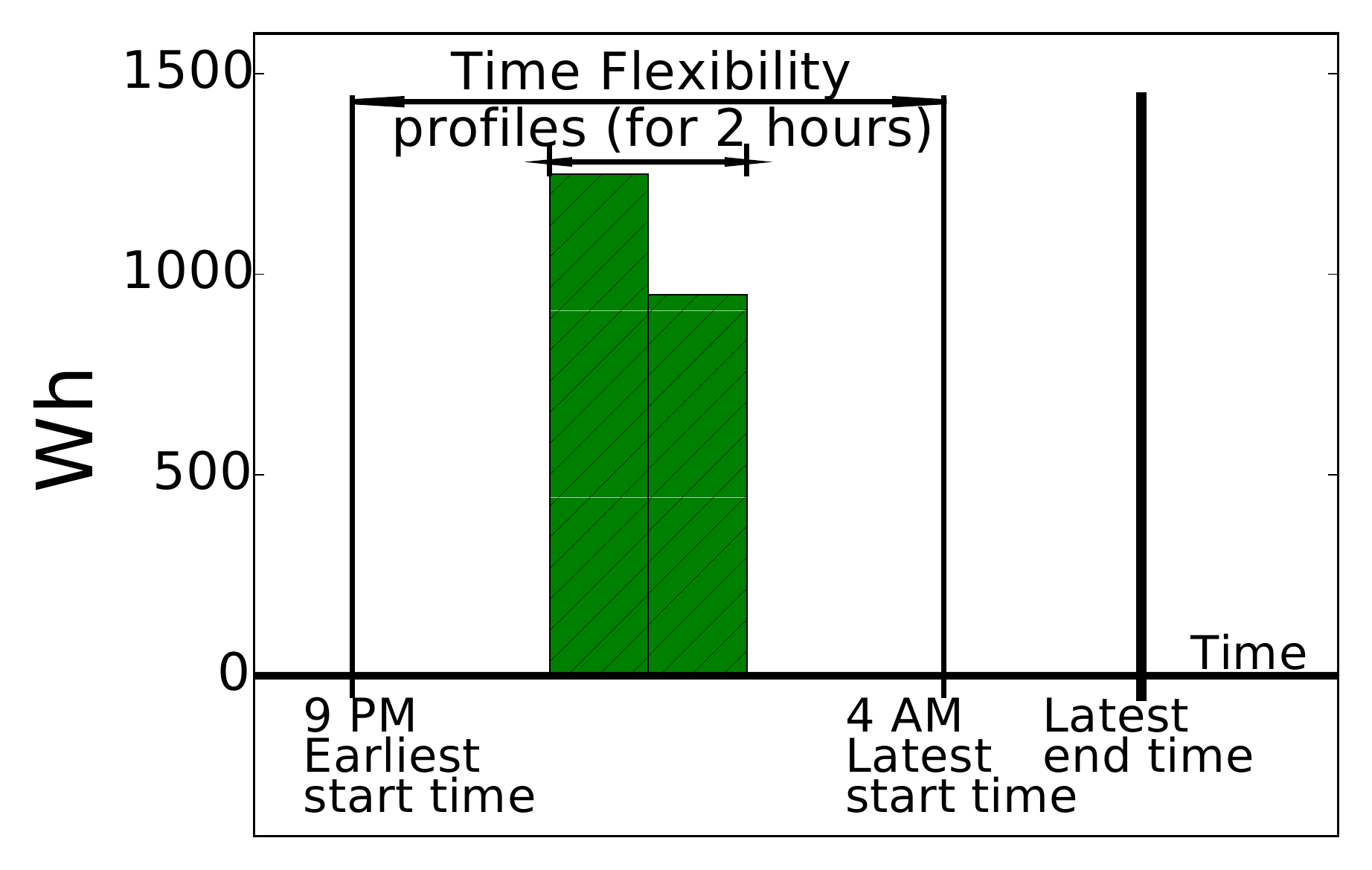} 
	\caption{A sample flex-offer for Washer Dryer.}
	\label{fig:Flex-Offer}
\end{figure}

The flexibilities extracted from individual devices such as EVs, heat pumps, washer dryers, PVs, etc., are generalized to generate the so-called \textit{micro flex-offers} \cite{Siksnys2015}. A single flex-offer includes:
\begin{itemize}
	\item An energy profile, representing energy demand for each discrete time units, e.g., per 15 minutes, of a operation.
	\item The time flexibility interval specifying a time duration during which device's operation can be preponed or postponed.
\end{itemize}

Modeling of flexibility from a variety of devices into a unified flex-offer object simplifies the aggregation and disaggregation across various dimensions, where no specific knowledge about the underlying device is needed. Figure \ref{fig:Flex-Offer} shows an example of a micro flex-offer generated from the extracted flexibility of a \textit{washer dryer}. The flex-offer in the figure states that the \textit{washer dryer} could be activated anytime between 9 PM and 4 AM and operates for 2 time units. It further shows the energy profile for the \textit{washer dryer} representing the demand for each time unit  of the devices' operation, and a constraint that once activated the \textit{washer dryer} should be operated continuously for 2 time units.

\begin{figure}
	\includegraphics[width=1\columnwidth, trim=0px 0px 0px 0px, clip=true]{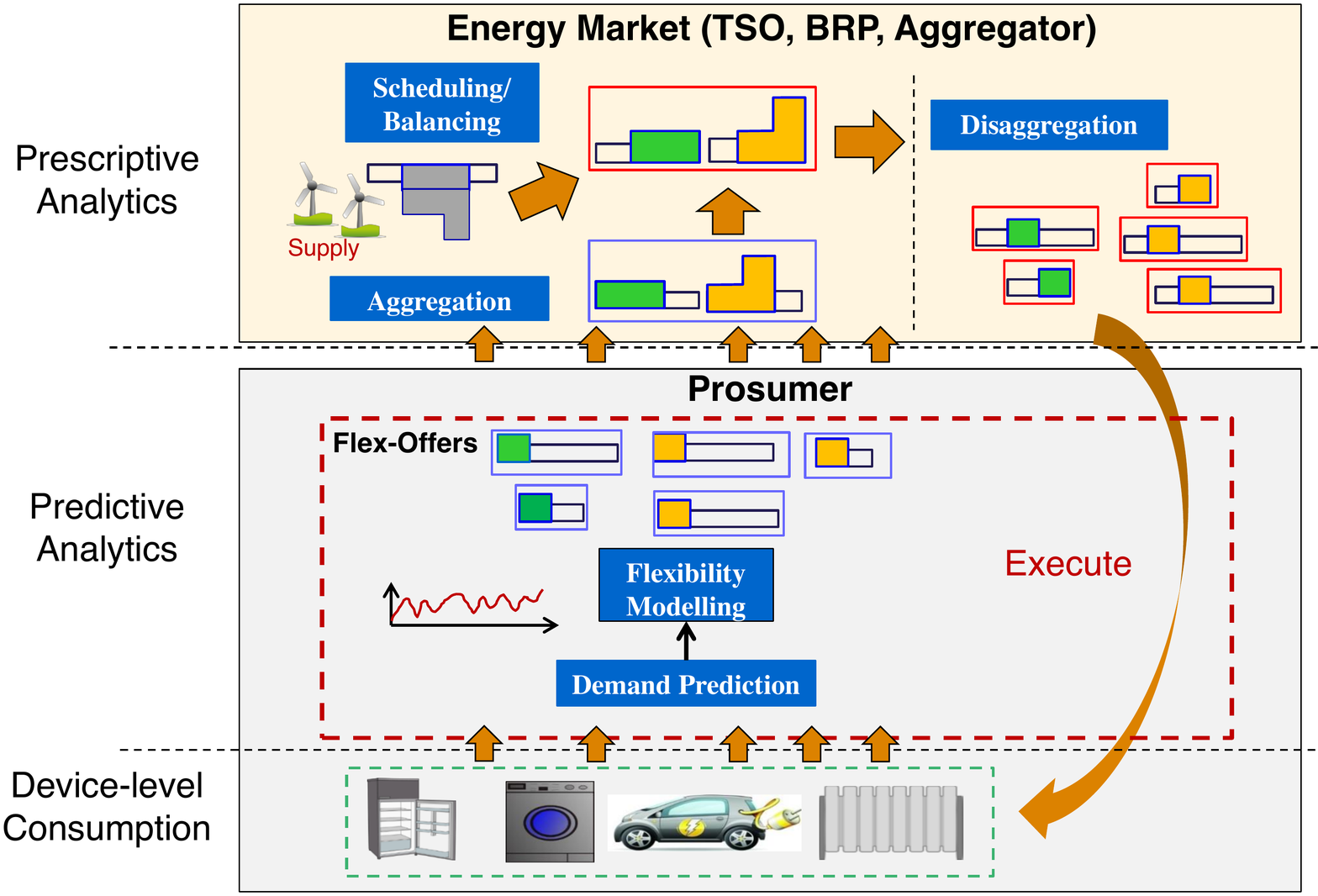} 
	\caption{The lifecycle of flex-offers, from generation to execution.}
	\label{fig:Flexlifecycle}
\end{figure}

Figure \ref{fig:Flexlifecycle} demonstrates the complete lifecycle of the flex-offers in the flexibility market. The concept is to analyze past consumption, usage patterns, operation correlations, and energy profiles of individual devices, and then forecast the available future demand flexibilities. The forecasted flexible demands are modeled as flex-offers. The device-level flex-offers typically have a small size and cannot be directly traded on the market. Hence, aggregators aggregate the flex-offers from individual devices into fewer, larger flex-offers, known as macro flex-offers. The flex-offers can be exploited by market players such as TSO, DSO, and Aggregator to balance demand and supply better or to delay costly grid upgrades. For example, a BRP schedules the flexible demands of macro flex-offers to obtain a global balance, i.e., to reduce the financial loss due to an error in demand forecast or intermittent RES production. Once the macro flex-offers have been scheduled (traded), the aggregators disaggregate the macro flex-offers into the respective schedules denoting the exact time and amount of energy that has to be consumed by each individual device. Finally, the flexible devices operate based on the received disaggregated schedules.

Here, the key assumption is that the operation of some of the devices can be automatically controlled and users are willing to provide flexibility as their contribution to the demand management in return for financial or other incentives. Depending on the size of the flex-offer, the control can be either performed directly by the existing market players such as BRP, DSO, or may be delegated to a new entity such as an Aggregator. This assumption is highly supported by the recent development in intelligent household appliances, where already today they can be set to run later, or response to energy price signals, e.g., washer dryer can be scheduled to operate later or operate a freezer at cost optimized mode. A user can always override the market proposed schedule, in such case, he will not receive the financial benefit for that flex-offer.

The flexibility-based DR scheme discussed above relies on the forecasted flexible demands of individual devices for flex-offer modeling, trading, and dynamic scheduling of demand and supply. Device-level demand forecasting is a challenging task due to highly stochastic user behavior. However, market players are more concerned with the utility that the forecasted flexibility brings to the market rather than the intrinsic quality of the forecast itself, such as a value of precision and recall. Hence, in this paper, we will evaluate the applicability of the frequently used forecast model for device-level demand and flexibility prediction. For simplicity, and to best analyze the effect of individual forecasts errors, we will assume that an individual micro flex-offer is traded in the market without any aggregation. In this paper, we analyze flexibility potential of household wet devices such as dishwasher, washer dryer, etc., which are typically operated once or twice a day. Further, these devices have minimal variation in power consumption within a given operation state, e.g., during the heating cycle. Hence, we will investigate time flexibility ranges of 1 to 24 hours, and the forecasted device-level demand represents the amount flexibility. In the next section, we will present the regulating market and discuss how demand flexibility can be used to save money in this market or avoid it altogether.

\section{Regulating Power Market}
The Nordic energy market plays an important role in balancing the supply and demand in the spot market for electricity in the Nordic countries. This regulating power market \cite{Frieder2011} is activated shortly before the time of the actual delivery and purchase of the power when the market is anticipated to have an imbalance in supply or demand. The regulating power could be activated for any duration of time. For our experiment, we assume that the duration of activation of regulating power is in unit of an hour. This assumption is not essential for our analysis and could be changed if a more fine-grained control should be desired. Regulating power can be either up or down as a consequence of the following situations. 

If the supply from a BRP deviates below its previous commitment to the spot market, the BRP has to buy up-regulating power -- at up-regulating power price -- in order to fulfill its commitment. The required amount of \textit{up-regulating} power is fulfilled by purchasing with other energy suppliers.
On the other hand, if the supply is greater than the previous commitment, the BRP has to sell \textit{down-regulating} power -- at down-regulating power price - or curtail the supply to maintain the energy balance in the market. The regulating power prices differ from the spot price. Thus the BRP suffers a financial loss when using the regulating power. With the introduction of flexibility-based demand response, a BRP can always schedule some portion of flexible demands in a way that maintain their portfolios and avoid regulating market.

Here, we define various parameters associated with the regulating power market.

\begin{itemize}
	\item Spot price, $p_s(t)$: Energy price at the spot market.
	\item Up-regulation volume, $v_u(t)$: The amount that is less than the actual demand in the spot market. 
	\item Down-regulation volume, $v_d(t)$: The amount that exceed the actual demand in the spot market.
	\item Up-regulating power price, $p_u(t)$: Price paid for the up-regulating power.
	\item Down-regulating power  price, $p_d(t)$: Price received for down-regulating power.
\end{itemize}
At any point in a time, one of the regulation volumes in the pair $(v_u(t), v_d(t))$ will be zero. For notational convenience, we will in the following represent the regulation volumes with a single notation.

Up/Down-regulation volume, $v_{u/d}(t)$: denotes the non-zero regulating element, or zero if both elements in the pair are zero.
\\
In the next section, we will discuss our implementation of logistic regression and pattern sequence matching algorithms for device-level demand forecasting.

\section{Device-level Forecasting}
Any device at a particular timestamp could be in one of the three possible states, i) \textit{idle}- switched off, ii) \textit{activation} - switched on, or  iii) \textit{operating}. The \textit{idle} and \textit{operating} state of a device are represented by 0 and activation state by 1. Further, a \textit{threshold value} represents the minimum power (watts) demand for a device to be in the \textit{activated} or \textit{operating state}. Device-level energy demand prediction for intermittently operating devices, such as dishwasher, washer dryer, etc.,  is not straightforward. Since they are in \textit{operating} state for the only couple of hours in a day and the rest of the time they are in the \textit{idle} state. Hence, in this paper, we predict device-level energy demand in two steps. First, we use the logistic regression model to predict device activation for each hour of a day and then use the pattern sequence matching to extract energy demand and operation duration for the predicted device activations.
\subsection{Dataset}

\begin{figure}
	\centering
	\includegraphics[width=.7\columnwidth, trim=5px 5px 5px 5px, clip=true]{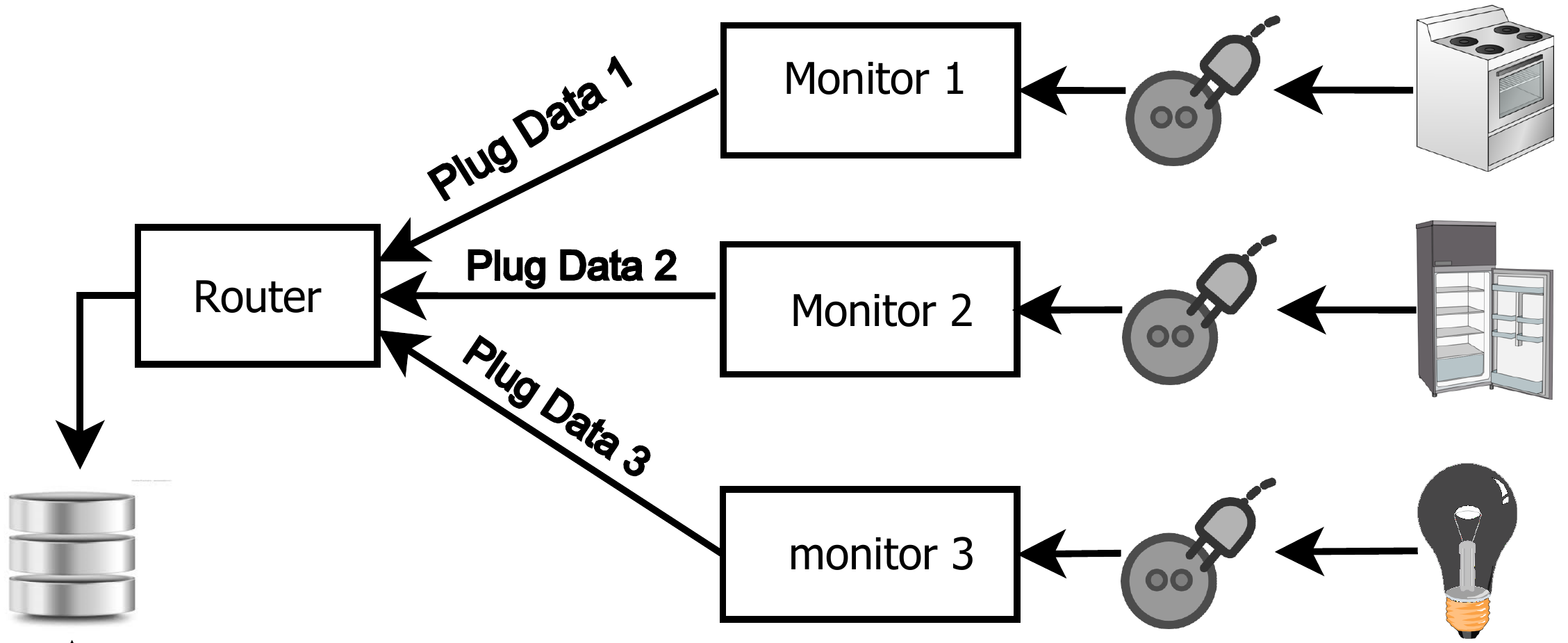} 
	\caption{The hardware architecture for device-level data collection.}
	\label{fig:plugdiagram}
\end{figure}

The device-level energy consumption dataset is collected using the smart plugs, as shown in Figure \ref{fig:plugdiagram}. 
The device-level dataset contains the average power readings in watts for individual devices.  
The dataset is logged at a frequency of once every 15 minutes and is collected through January 2014 to October 2015. We utilize the first 80\% of the timeseries as the training and validation set and the remaining 20\% as the test set.
To evaluate the financial implications that are caused by an error in demand forecast, we use the energy market dataset from the Danish TSO Energynet.dk.  The dataset consists of hourly energy prices in the spot and regulation market, as shown in Table \ref{tbl:Data}. 

\begin{table}
	\begin{adjustbox}{max width=.48\textwidth}
		\centering
		\begin{tabular}{|c|c|c|c|c|c|c|}
			\hline
			Date     & Hour & \begin{tabular}{@{}c@{}}Up-regulation \\ Volume\end{tabular} & \begin{tabular}{@{}c@{}}Down-regulation \\ Volume\end{tabular} &\begin{tabular}{@{}c@{}}Up-regulation \\ Price\end{tabular} & \begin{tabular}{@{}c@{}}Down-Regulation \\ price\end{tabular} & \begin{tabular}{@{}c@{}}spot \\ price\end{tabular} \\ \hline
			1/1/2016 & 0    & 200                   & 0                      & 222.43               & 113                   & 113        \\ 
			-        & -    & -                     & -                      & -                    & -                     & -          \\ 
			-        & -    & -                     & -                      & -                    & -                     & -          \\
			1/1/2016 & 9    & 0                     & 0                      & 189.84               & 137.91                & 189.84     \\ 
			1/1/2016 & 10   & 0                     & -84                    & 183.71               & 164.6                 & 183.71     \\
			\hline
		\end{tabular}
	\end{adjustbox}
	\caption{Market Price Data}
	\label{tbl:Data}
\end{table}

\subsection{Data Resolution}
The stochasticity associated with device-level demand makes forecasting a difficult task. Further, the operation of a device depends on various external factors and predicting the device state at, e.g., an hourly resolution is challenging in the absence of context information. Hence, in this paper, we take a more general approach. We assess the accuracy of typical forecast models at different data granularities and evaluate the one best suited for flexibility-based DR in terms of the utility it provides for the market.

We assume a time series dataset $X = \{a_1, a_2, \dots, a_{d-1}\}$ of device activation profiles for $d-1$ days,
where each $a_i$ value depends on the aggregation level discussed below. 
We will consider aggregation into the 3 most commonly analyzed data granularities.

\textbf{Hourly Resolution:} Here, the forecasting problem is to predict the hourly device activation probability for the next 24 hours. Thus, the device-level consumption data is aggregated to an hourly resolution with the energy consumption replaced by a binary activation value. 
Specifically, if a reading is above a \textit{threshold value} and represents an activation state then the reading is replaced by 1 else by 0. 
In this case, the dataset X represents an hourly device activation profile. Hence, each $a_i \in \{0,1\}^{24}$ is a vector composed of 24 hourly profiles corresponding to certain day $i$.

\textbf{Group Resolution:} Here, the forecasting problem is to predict the device activation probability for each group in the next 24 hours. 
The groups are created by clustering the 24 hours into $m$ groups based on the operational probability at each hour, e.g., we create a set of group $G = \{g_1, g_2, g_3\},$ where $g_1 = \{1,\dots,7\}, g_2 = \{8,\dots,15\}$, and $g_3 = \{16,\dots,24\}$. 
The hourly dataset is here aggregated to the group resolution. Specifically, if any hour in the group has value 1, then the group gets value 1 otherwise it is set to value 0. In this case, the dataset X represents the device activation profile for the groups. Hence, each $a_i \in \{0,1\}^m$ is a vector composed of activation profile for $m$ groups corresponding to a certain day $i$.

\textbf{Daily Resolution:} Here, the forecasting problem is to predict the probability of the device activation in the following day. Thus, the dataset is here aggregated to the daily resolution. Specifically, if any single reading in the day is greater than the \textit{threshold value}, then the day gets value 1 otherwise it is set to 0. In this case, the dataset X represents daily device activation profile and each $a_i \in \{0,1\}$ is the activation state corresponding to a certain day $i$.

\subsection{Feature Extraction}
We analyse the collected device-level dataset with an aim to extract features that can reliably capture the \DAP{} and energy demand. More specifically, we generate additional derived values from the initial measured data to enhance the information on the device activation and usage patterns. The descriptions of some of the extracted features is described in the Table \ref{tbl_features}.  

The present state of a device is highly dependent on its previous states, i.e., a device with no recent activities has a higher probability of activation than the devices recently activated. 
Thus, we extract the device states in the \textit{previous 24 hours} as 24 binary features and an additional 7 binary features to represent the time since the \textit{last operation} (1 and 6 in Table \ref{tbl_features}).  For the daily forecast, we extract the device activation patterns for the past $w$ days, where $w$ is the window size (2 in Table \ref{tbl_features}). We assume that the uses of devices have some temporal patterns, e.g., an oven is mostly activated during the morning and evening, and the dishwasher is mostly operated after the lunch or dinner, etc. Further, we can notice a variation in device activation patterns during the days of the week. Therefore, we generate 24 binary features to represent each\textit{ hour of the day} and 7 binary features representing the \textit{day of the week} (3 and 4 in Table \ref{tbl_features}).

To capture the influence of seasonal factors on the usage patterns, we include four binary features representing the four \textit{seasons} of the year (7 in Table \ref{tbl_features})). In addition, we create various additional features as a multiplicative interaction between the above-extracted features. We will in the following use $x_i = {x_i^1, x_i^2,  \dots, x_i^m} $  to represent m features corresponding to a data point $i$ in $X$, and use  the convention that $x^m$ refers to the $m^{th}$ feature and $x^{\{m\}}$ refers to a set with $m$ features. 
\begin{table*}[]
	\centering
	\begin{adjustbox}{max width=.94\textwidth}
		\label{my-label}
		\begin{tabular}{llll|l|c|c|c|}
			\cline{6-8}
			&                                                         & &                                                                                                                                                                                                                                                                                                                                                             &        & \multicolumn{3}{c|}{\textbf{Feature Used}}                                                                                                                \\ \hline
			
			\multicolumn{1}{|c|}{\textbf{S.No.}} &\multicolumn{1}{|c|}{\textbf{Features}} & \multicolumn{1}{l|}{\textbf{Notation}} & \multicolumn{1}{c|}{\textbf{Description}}                                                                                                                                                                                                                                                                                                                       &\multicolumn{1}{c|}{\textbf{Example}} & \multicolumn{1}{c|}{\textbf{Hourly}}                                          & \multicolumn{1}{l|}{\textbf{Group}} & \multicolumn{1}{l|}{\textbf{Daily}} \\ \hline 
			%
			
			\multicolumn{1}{|c|}{1}&\multicolumn{1}{|l|}{Last 24 States, $L(x^{\{24\}})$}                     & \multicolumn{1}{l|}{if $h = 0$ then \{$a_{i-1}^0, \dots, a_{i-1}^{23} $\} else \{$a_{i-1}^h, \dots, a_i^{h-1} $\}}                                   & 24 Binary Features                                                                                                                                                                                                                                                                                                                                          & {$\{0,0,1, \dots, 0, 1 \}$} &X                                                                          & X                                   &                                     \\ \hline
			
			\multicolumn{1}{|c|}{2}&\multicolumn{1}{|l|}{Last 7 States, $L(x^{\{7\}})$}                      & \multicolumn{1}{l|}{\{$a_{i-7}, \dots, a_{i-1}$\}}                                   & 7 Binary Features                                                                                                                                                                                                                                                                                                                                           & {$\{1,0,1,1,0,1,1\}$} &                                                                              &                                     & X                                   \\ \hline
			
			\multicolumn{1}{|c|}{3}&\multicolumn{1}{|l|}{Hour of a Day $H(x^{\{24\}})$}                      & \multicolumn{1}{l|}{$\chi^{hour}(x_i) = \begin{cases}
				1 & \text{ if } hour \ of\ the\ prediction \ point\\ 
				0 & \text{ otherwise}  
				\end{cases}$}                                   & \begin{tabular}[c]{@{}l@{}}24 Binary Features\\ hour $\in \{0, \dots, 23\}$\end{tabular}                                                                                                                                                                                                                                                                & {\begin{tabular}[c]{@{}l@{}}For h = 2\\ $\{0,0,1,0,\dots, 0, 0 \} $\end{tabular}} &X                                                                        & X                                   &                                    \\ \hline
			
			\multicolumn{1}{|c|}{4}&\multicolumn{1}{|l|}{Days of the Week, $D(x^{\{7\}})$}                   & \multicolumn{1}{l|}{$\chi^{day}(x_i) = \begin{cases}
				1 & \text{ if } the \ prediction \ day\\ 
				0 & \text{ otherwise}  
				\end{cases}$}                                   & \begin{tabular}[c]{@{}l@{}}7 Binary Features\\ day $\in \{mon, \dots, sun\}$\end{tabular}                                                                                                                                                                                                                                                                   & {\begin{tabular}[c]{@{}l@{}}For Thursday \\ $\{0,0,0,1,0,0,0\} $\end{tabular}} &X                                                                        & X                                   &                 X                    \\ \hline
			
			\multicolumn{1}{|c|}{5}&\multicolumn{1}{|l|}{Is weekend, $W(x^{\{1\}})$}                         & \multicolumn{1}{l|}{$\chi(x_i) = \begin{cases}
				1 & \text{ if } weekend \\ 
				0 & \text{ otherwise}  
				\end{cases}$}                                   & 1 Binary Feature & {\begin{tabular}[c]{@{}l@{}}If $sat,\ or\ sun$ than 1 \\ else 0\end{tabular}} &X                                       & X                                   & X                                   \\ \hline
			
			\multicolumn{1}{|c|}{6}&\multicolumn{1}{|l|}{Last Operation, $LO(x^{\{7\}})$}                     & \multicolumn{1}{l|}{$\chi^{n}(x_{i-7}, \dots, x_{i-1}) = \begin{cases}
				1 & \text{ if } \ x_{i-n} =1 \ and \ \forall \{x_{i-n+1}, \dots, x_{i-1}\} = 0\\ 
				0 & \text{ otherwise}  
				\end{cases}$}                                   & \begin{tabular}[c]{@{}l@{}}7 Binary Features \\ n $ \in \{1,2,3,4,5,6,\geq 7\}$ \end{tabular} & {\begin{tabular}[c]{@{}l@{}}If last operation was \\ 2 days before, than\\ $\{0,0,0,0,0,1,0\} $ \end{tabular}} &X                                                                         & X                                   & X                                   \\ \hline
			
			\multicolumn{1}{|c|}{7}&\multicolumn{1}{|l|}{Season, $S(x^{\{4\}})$}                     & \multicolumn{1}{l|}{$\chi^{s}(x_{i}) = \begin{cases}
				1 & \text{ if } season \ of\ the\ prediction \ point \\  
				0 & \text{ otherwise}  
				\end{cases}$}                                   & \begin{tabular}[c]{@{}l@{}}4 Binary Features \\ s $ \in \{winter, spring,$ \\ $ summer , autumn\}$ \end{tabular} & {\begin{tabular}[c]{@{}l@{}}If spring, than\\ $\{0,1,0,0\} $ \end{tabular}} &X                                                                        & X                                   & X                                   \\ \hline
			
		\end{tabular}
	\end{adjustbox}
	\caption{Features Description}
	\label{tbl_features}
\end{table*}

\subsection{Filling Observation Gaps}
The device-level dataset used for this experiment contains observation gaps that has to be filled to maintain the continuity of the time series. There exist many alternative methods suggested for imputing these missing values. However, the device-level datasets are usually skewed with many instances of inactive state and much fewer instances of the activation state. 
Thus, in this experiment we replace all observation gaps with the value of inactive state, i.e., 0. Further, we include an additional binary "is missing" feature for each data point used during feature generation. For example, for the features set representing the last 24 hours readings, any missing values are set to 0 and "is missing" features for those points are set to 1.

\subsection{Device Activation and Demand Forecasting}
This subsection details our forecast model and approach for device-level demand prediction. The Logistic Regression model is used to predict a device activation time, and a pattern matching approach is used to estimate operation duration and energy demand for the predicted activation.

\textbf{Logistic Regression (LR): }
The standard logistic regression \cite{hosmer2004} model has been used extensively in the literature for various binary classification problems \cite{chatterjee2015, SALASELJATIB2018}. The model defines the relationship between a set of explanatory variables and a dependent classification variable, and provides the probability or likelihoods of the possible outcomes. Let $Y = \{y_1, y_2, \dots, y_n\}$ be binary dependent variables where each $y_i \in \{0,1\}$ represents the class label for the feature vector $x_i$. Let, $z_i$ represent a linear function of the explanatory variables:
\begin{center}
	$z_i = \theta^0 + \sum_{j=1}^{m}\theta^j x_i^j$,
\end{center}
where $\theta = (\theta^1, \theta^2, \dots, \theta^m)$ are the regression parameters associated with the explanatory feature vector $x_i = (x_i^1, x_i^2, \dots,$ $ x_i^m)$
The probability that a new data point belongs to the class label 1 is represented as:
\begin{center}
	$p(y_i=1|x_i;\theta) =  \frac{1}{1+e^{-z_i}}$,\\
\end{center}
Let us use $\pi_i = p(y_i=1|x_i;\theta) $  to simplify the notation, the probability of the possible outcome to be class 0 is represented as:   $p(y_i=0|x_i;\theta) = 1 - \pi_i $ .
%
Given the training data $((x_1,y_1), \dots,  (x_n,y_n))$, the optimal regression coefficients $\theta$ can be estimated by maximizing  the log-likelihood function:
\begin{center}
	$L(\theta) = \sum_{n}(y_n \  ln\pi_n + (1-y_n) ln(1-\pi_n))$,
\end{center}
We are considering a high number of features in the model and therefore introduce L1 regularization in order to counter overfitting the model to the training data. 
The L1 regularized log-likelihood function is:
\begin{center}
	$L(\theta) = \sum_{n}(y_n \  ln\pi_n + (1-y_n) ln(1-\pi_n))  + \lambda \sum_{j=1}^{m}|\theta^j|$
\end{center}
where $\lambda$ is the regularization parameter. 

\textbf{Demand Forecast: }

After the prediction of device activation time, we use the Pattern Sequence Matching (PSM) algorithm to estimate the duration of the device operation and the energy demand for each time unit of operation. Pattern Sequence Matching works under the premise that the energy consumption patterns and operation durations are correlated to the hour of device activation, e.g., a washer dryer activated at 17:00 always operates for two time units and has an average energy profile of <1.6, 1.1> kWh.

\begin{figure}
	\centering
	\includegraphics[scale=.45]{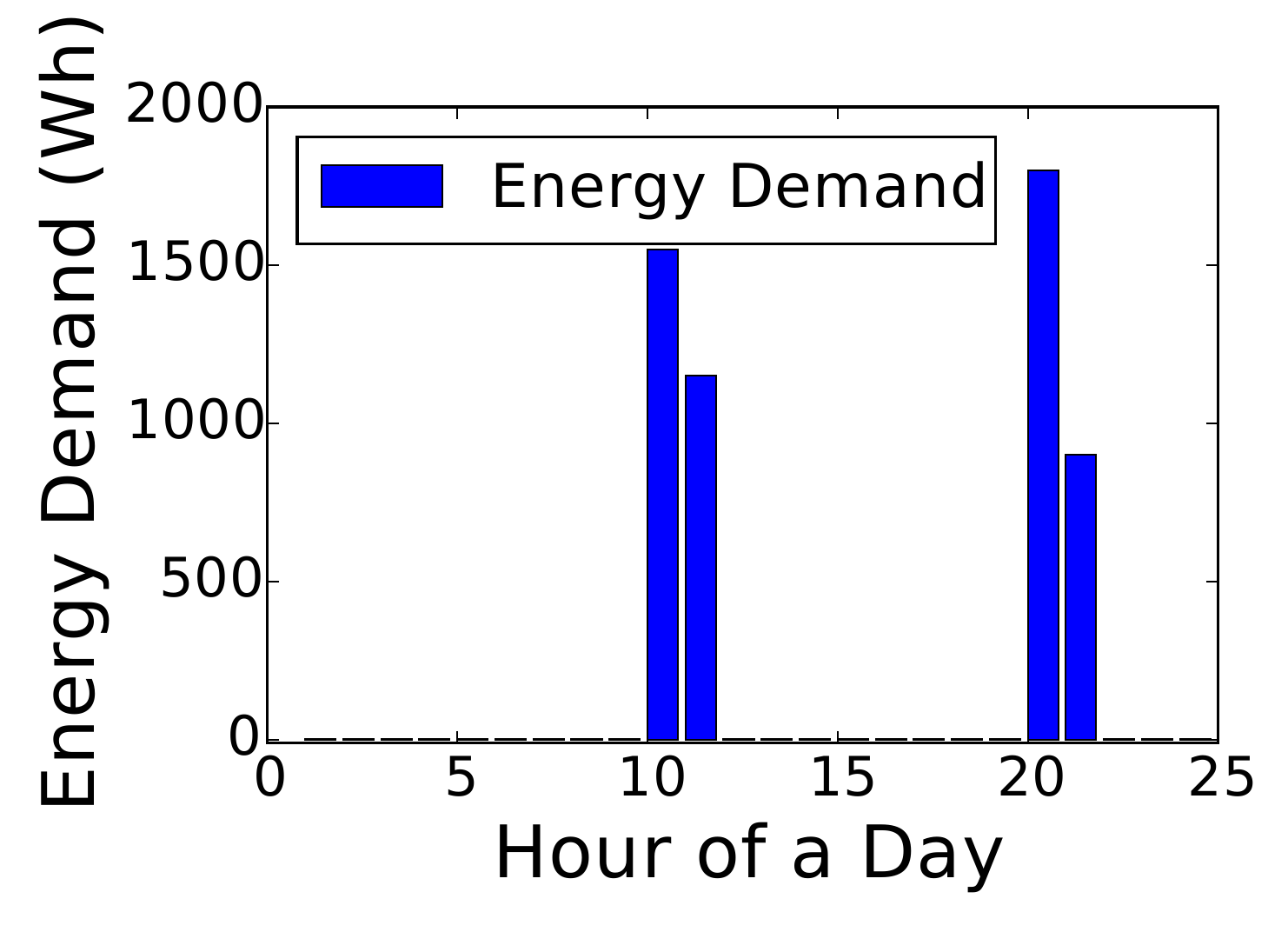} 
	\caption{Demand Forecast - for next 24 hours}
	\label{fig:forecast}
\end{figure}

Therefore, to estimate the energy profile for a predicted device activation at hour $h$ of day $d$, the PSM first searches device activations triggered at hour $h$ in the historical time series X. Then, for each activation the algorithm extracts the energy demand $\{a_i^h, \dots, a_i^{(h+k-1)}\}$ for $k$ duration of the device operation. This search outputs device activation profiles $P = \left \langle p_1, \dots, p_n \right \rangle$, where each $p_i$ is an energy profile of the device activation and $n$  is the number of activations at the hour $h$. 
Next, the operation duration $l$ for the forecasted device activation is estimated as the ceiling of the average operation duration in the historical profiles $P$,  $\mathit{l = \left \lceil \frac{1}{n} \sum_{i=1}^{n} |p_i| \right \rceil}$, where $p_i \in P$. Then, the energy demand for each time unit of operation is calculated as the average energy consumption at the respective time unit in the pattern p, detailed in Algorithm 1. If there are no device activations at hour $h$ in $X$, i.e., $P \leftarrow \emptyset$, then the energy profile is extracted utilizing all historical device activations. Figure \ref{fig:forecast} shows an example of demand forecast at hourly granularity, with 2 predicted device activation for the day $d$.

\begin{algorithm}[h]
	\caption{Pattern Sequence Matching (PSM): Hourly}
	\begin{algorithmic}[1]
		\State \textbf{Input:} =  $X$ - $\{a_1,\dots,a_{d-1}\}$ a time series.
		\State \qquad \qquad $h$ -  a predicted device activation hour.
		\State \textbf{Output:} = $p$ - energy profile.
		\Function{hourlyDemandForecast}{$X, h$}
		\State \textit{P} $ \leftarrow \emptyset $; $ p \leftarrow \emptyset $; $ l \leftarrow 0$
		\For {$i \leftarrow 1:d-1$}
		\For{$h \leftarrow 1:24$}
		\If{$a_i^h \geq thres$ and $a_i^{h-1} == 0$}
		\While{$a_i^h \geq thres$}
		\State $p \leftarrow p \cup \{a_i^h\}$;
		\If{$h == 24$}
		\State $h \leftarrow 0$;$i \leftarrow i+1$
		\Else
		\State  $h \leftarrow h+1$	
		\EndIf	 
		\EndWhile
		\EndIf
		\State $P \leftarrow P \cup \{p\}$ ; $  p \leftarrow \emptyset$ ; $ l \leftarrow l+ |p|$
		\EndFor
		\EndFor
		\State $\mathit{l = \lceil \frac{l}{|P|} \rceil}$
		\State $ p \leftarrow \emptyset$
		\For {$j \leftarrow 1:l$}
		\State $s_j \leftarrow \frac{1}{n}\sum_{i=1}^{n}P_i.a_j$
		\State $p \leftarrow p \cup \{s_j\} $
		\EndFor
		\State Return $p$
		\EndFunction
		\label{PSM}
	\end{algorithmic}
\end{algorithm}

To estimate the energy profile at group resolution, we follow the same procedure with a dataset X which is instead at group resolution and $h$ replaced with $g$. 
Similarly, to estimate the energy profile at daily resolution, we extract patterns for all activations at the day of the week $w$ in $X$, where $w$ is the day of the week for the day $d$, described in Algo 2.

\begin{algorithm}
	\caption{Pattern Sequence Matching (PSM): Daily}
	\begin{algorithmic}[1]
		\State \textbf{Input:} daily resolution dataset $X= \{a_1,\dots,a_{d-1}\}$. 
		\State \textbf{Output:}  $p$ - energy profile.
		\Function{DailyDemandForecast}{$X$}
		\State \textit{P} $ \leftarrow \emptyset $
		\For {$i \leftarrow 1:d-1$}
		\State $ p \leftarrow \emptyset $; $ l \leftarrow 0$ ; $active \leftarrow false$
		\If{$dayofweek(a_i) == dayofweek(a_d)$}
		\For{$h \leftarrow 1:24$}
		\If {$a_i^h \geq thres$}
		\State $p \leftarrow p \cup \{a_i^h\}$; $active \leftarrow true$
		\Else 
		\If {$active = true$}
		\State $P \leftarrow P \cup \{p\}$; $ l \leftarrow l+ |p|$
		\EndIf
		\State $p \leftarrow \emptyset$ ; $active \leftarrow false$ 
		\EndIf
		\EndFor
		\EndIf
		\EndFor
		\State $\mathit{l = \lceil \frac{l}{ |P|} \rceil}$
		\State $ p \leftarrow \emptyset$
		\For {$j \leftarrow 1:l$}
		\State $s_j \leftarrow \frac{1}{n}\sum_{i=1}^{n}P_i.a_j$
		\State $p \leftarrow p \cup \{s_j\} $
		\EndFor
		\State Return $p$
		\EndFunction
	\end{algorithmic}
\end{algorithm}

\subsection{Model Evaluation}
Precision, Recall, and Receiver Operator Characteristics (ROC) are commonly used in the literature for binary decision problems. However, for the class imbalanced dataset, the ROC curve does not provide the real picture of the performance of the model due to the slower increasing rate of the dominant class (96.9\% of instances in our dataset), i.e., false positive rate.  Therefore, in our experiment, we evaluate the performance of the classifiers on Area Under the precision-recall curve (PR-curve) 
as discussed in \cite{Davis2006}.
\begin{table}[h]
	\centering
	\begin{adjustbox}{max width=.44\textwidth}
		\begin{tabular}{|c|c|c|c|c|}
			\hline
			Actual  $f(t)$ & Predicted  $\hat{f}(t)$ & Type               & Benefit                   & Loss                         \\ \hline
			0             & 0                      & True Negative (TN) & 	  &                             \\ \hline
			0             & $>$ 0                    & False Positive (FP) &        $X$                  & $X$ \\ \hline
			$>$ 0           & 0                      & False Negative (FN)   & $X$  & $X$ \\ \hline
			$>$ 0           & $>$0                     & True Positive (TP) & $X$  &                             \\
			\hline
		\end{tabular}
	\end{adjustbox}
	\caption{Categories of forecast result: based on actual and forecasted demand.\label{tbl_confdetail}}
\end{table} 

Let $\hat{F} = \{\hat{f}(1), \hat{f}(2), \dots, \hat{f}(24)\}$ be 24 hourly forecasted demands for the day $d$, and $F = \{f(1), \allowbreak f(2),\dots, $ $f(24)\}$ be the actual demands at the time of delivery. 
Based on the forecasted and actual demand the results of a forecast model can be divided into four categories, shown in Table \ref{tbl_confdetail}. 
The table further shows the consequences, in terms of benefit and loss, that a market experience for each category of the result.  This individual analysis leverages the market players in selecting a forecast model with the best performance on the desired category. For example, a market can desire a model with a higher precision or a higher recall, or have a trade-off between precision and recall. 
In the next sub-section, we will discuss the impact of each category of the result on the flexibility-based demand response market.

\subsection{Impact of Forecast Result on Market}
For a flexibility-based DR market, a precise estimate of future demand and associate flexibility is desired for dynamic pre-scheduling of demand and supply. However, at the device-level, the demand at a particular time highly depends on various factors, such as user availability, preference, weather condition, device settings, etc. A forecast model suffers from stochastic user behaviors and external factors that are hard to capture, resulting in a higher forecast error. This may create a higher imbalance in the market due to the scheduling of false and unplanned energy demand. Therefore, a market player is always interested to know the maximum limit of forecast error that can be handled in the flexibility market without any further financial loss. In this regard, we will analyze the effect of each type of forecast result described in Table \ref{tbl_confdetail}.

For the \textit{TN} results ($f(h)$ = $\hat{f}(h) = 0$), the market neither has flexible demands to schedule nor experience any unexpected demands at the time of actual delivery. Thus, no financial loss or benefit comes with the \textit{TN}.  In the case of \textit{FP} results ($f(h) $ = 0 and $\hat{f}(h) > 0$), the loss depends on the market balance at the time of actual deliveries. For example, the up-regulated market at the scheduled timestamp can achieve financial gain by a reduction in regulation volume. 
On the other hand, the \textit{FP} will increase the anticipated total demand due to inaccurate estimation, which in turn causes the financial loss due to the change in the market prices, discussed in Section 5.3. 
Similarly, in the case of \textit{FN} results ($f(h)  > 0$  and $\hat{f}(h) = 0$),  an unscheduled demand could increase the up-regulation volume causing financial loss or decrease the down-regulation volume generating financial benefits. Finally, for the \textit{TP} results, the market generates financial benefit by pre-scheduling the flexible demand to reduce the regulation cost.
One can argue that aggregation could result in a mutual counterbalance of false positive and false negative at the same time-slot. However, we emphasize that depending on the market optimization objective; these two predictions might end up in two different aggregators (Fig. \ref{fig:Flexlifecycle}), thus providing no counterbalance effect. 
In the next section, we will quantify the benefit that can be achieved by shifting of the flexible demand and the corresponding loss due to the forecast error.

\section{Financial Evaluation}
In this section, we will define the savings that can be generated from the energy flexibility and analyze it in relation to the forecast error.

\subsection{Scheduling of Flexible Demand}
The extent to which a forecasted demand, i.e., flexible demand can be shifted is constrained by the time flexibility associated with the demand. Let $\tau \in \{0,1,2,\dots, $ $24\}$ be time flexibility associated with each forecasted demand in $\hat{F}$, where, in particular $\tau =0 $ corresponds to inflexible demands. Now, $\hat{F}=\{(\hat{f}(1),\tau_1), (\hat{f}(2),\tau_2), \dots,$ $ (\hat{f}{(24)},\tau_{24}) \}$ represents a vector of tuples, where $\tau_i$ is the time flexibility for $\hat{f}(i)$. To ease notation, we will assume the same fixed time flexibility for all the demands, but this assumption is easily generalized to varying time flexibilities across demands. 
Let, $V = \{v_{u/d}(1), v_{u/d}(2),\allowbreak  \dots,v_{u/d}(k)\}$ represent regulation volumes for the next $k$ hours from the time of forecast, where $k = {24+\tau}$ and $v_{u/d}(i)$ denotes the nonzero element of regulating volume, i.e., up- or down-regulation. Now, the scheduling task is to assign the flexible demands $\hat{F}$ to $V$ such that the market maximizes the total reduction in the regulation volume. Then, the optimization problem becomes:
\begin{equation*}
\begin{aligned}
& \text{maximize}
& & \sum_{i=1}^{k}{|v_{u/d}(i)| - \overline{|v_{u/d}(i)|}}  \\
& \text{subject to}
& & s_i \geq i,s_i \leq i + \tau, v_{u/d}(i)> \hat{f}(i) \\
\end{aligned}
\end{equation*}
where $s_i$ is the scheduled time for $\hat{f}(i)$ and the overbar in $\overline{|v_{u/d}(i)|}$ denotes the change in regulation volume due to shifting of $\hat{f}(i)$. To solve the optimization problem, we use the GLPK solver with PuLP in Python. The computational complexity of the optimization problem is $\mathcal{O}(n*k^l)$, where n is the number of activations for a device in next 24 hours and $l$ is the maximum operation duration. Since m, n, and d are relatively small integer values, on a standard laptop with 8GB RAM, 256GB SDD, and Intel Core i7 CPU with 4 Cores, the worst case running time for the solver is $<$ 3ms.
\subsection{Change in Regulation Price}
The inaccurate estimation of demand changes the anticipated regulation volume. Since the regulation prices in the market depend on the volume and type of regulation \cite{Skytte1999}, the change in volume affects the regulating power prices in the market. 
In the actual flexibility based market, flexibilities from small devices are aggregated into larger units and will have a bigger impact on the market. However, to estimate the impact of the forecast error at an atomic (device) level, we proportionate the aggregated effect to an individual device. 
Therefore, to evaluate the change in regulation price we use the hypothetical relationship between energy prices and regulation volume as proposed in \cite{Neupane2015}. 
\begin{align}
p_{u/d}(i)& = 1 \cdot p_{s}(i)  \nonumber \\
& + 1_{{v_d(i)}{<0}}(-0.334 \cdot p_s(i) +.0005 \cdot (p_s(i) \cdot v_d(i)) )\nonumber\\ 
& + 1_{{v_u(i)}{>0}}(.238  \cdot p_s(i) +.0034 \cdot (p_s(i) \cdot v_u(i)) ) \label{eq:3}
\end{align}
Here, $1_{a<b}$ denotes the indicator function for the predicate $a < b$, and $ p_{u/d}(i)$  is the predicted up-regulating power price $p_u(i)$ in case of up-regulation the predicted down-regulating power price $p_d(i)$ in case of down-regulation.
\subsection{Savings in Regulation Cost}
For each hour in V, the loss due to the market imbalance is computed as a product of the regulation volume times the price difference between regulating and the spot price. Hence, the total regulation cost for $V$ is calculated as:
\begin{align}
R=\sum_{i=1}^{k}v_{u/d}(i) \cdot |p_{u/d}(i) - p_s(i)|
\label{eqn_rcost}
\end{align}
where $p_{u/d}(i)$ is regulation price and $p_s(i)$ is spot price. Given, the regulation volume and forecasted flexible demand $\hat{F} $, the market generate a demand schedule that minimizes the regulation volumes. Let the new expected regulation volumes be $\bar{V}| \{ \forall i, \overline{v_{u/d}(i)} \leq v_{u/d}(i)\}$. Hence, the total expected regulation cost $E$ is given by: 
\begin{align}
E = \sum_{i=1}^{k}\overline{v_{u/d}}(i) \cdot |p_{u/d}(i) - p_s(i)|
\end{align}

The objective of scheduling the flexible demands is to reduce the regulation cost of the market. Thus, the expected regulation cost $E$ is always less than or equal to $R$, i.e., $E \leq R$. Therefore, savings in regulation cost due to shifting of flexible demand is given by: $\Delta R = R - E$

At the time of actual delivery, if the demand deviates from the previously forecasted demand the market player, i.e., BRP that caused the specific imbalance is financially responsible for the deviation. The total financial loss due to the error in demand forecast is calculated as:
\begin{align}
L   & = \sum_{i=1}^{k}|f(i) - \hat{f}(i)| \cdot |p_{u/d}(i) - p_s(i)| \nonumber \\
&+ 1_{f(i) \neq \hat{f}(i)}(v_{u/d}(i) \cdot |\overline{p_{u/d}(i)} - p_s(i)|)
\end{align}
where overbar denotes the updated regulation price calculated using the update regulation volume, i.e., $\overline{v_{u/d}(i)} = v_{u/d}(i) \pm f(i) $, and $|f(i) - \hat{f}(i)|$ is the difference between the actual and forecasted demand.
Recall the classification of forecast results in Table \ref{tbl_confdetail}, $\Delta R$ and $L$ represents the benefit and the loss for the table, respectively. The $\Delta R - L $ gives the total benefits of shifting flexible demands. One could argue that the financial loss can come from error in time flexibility. However, in this paper, we perform the cost-benefit analysis for a wide range of possible time flexibilities, i.e., 1 to 24 hours, instead of some particular value. Thus, the cost-benefit analysis holds even without considering the effect of time flexibility.

\section{Experimental Evaluation}
We perform a number of experiments, utilizing real-world demand measurements and market data, to analyze the viability of introducing our flexibility-based DR scheme in a energy market. In this section, we quantify the savings in regulation cost that can be achieved by utilizing device-level demand flexibility and then analyze the savings in relation to forecast accuracy.

\subsection{Savings in the Regulation Market}
\begin{figure}
	\centering
	\includegraphics[width=.9\columnwidth, trim=0px 0px 0px 0px, clip=true]{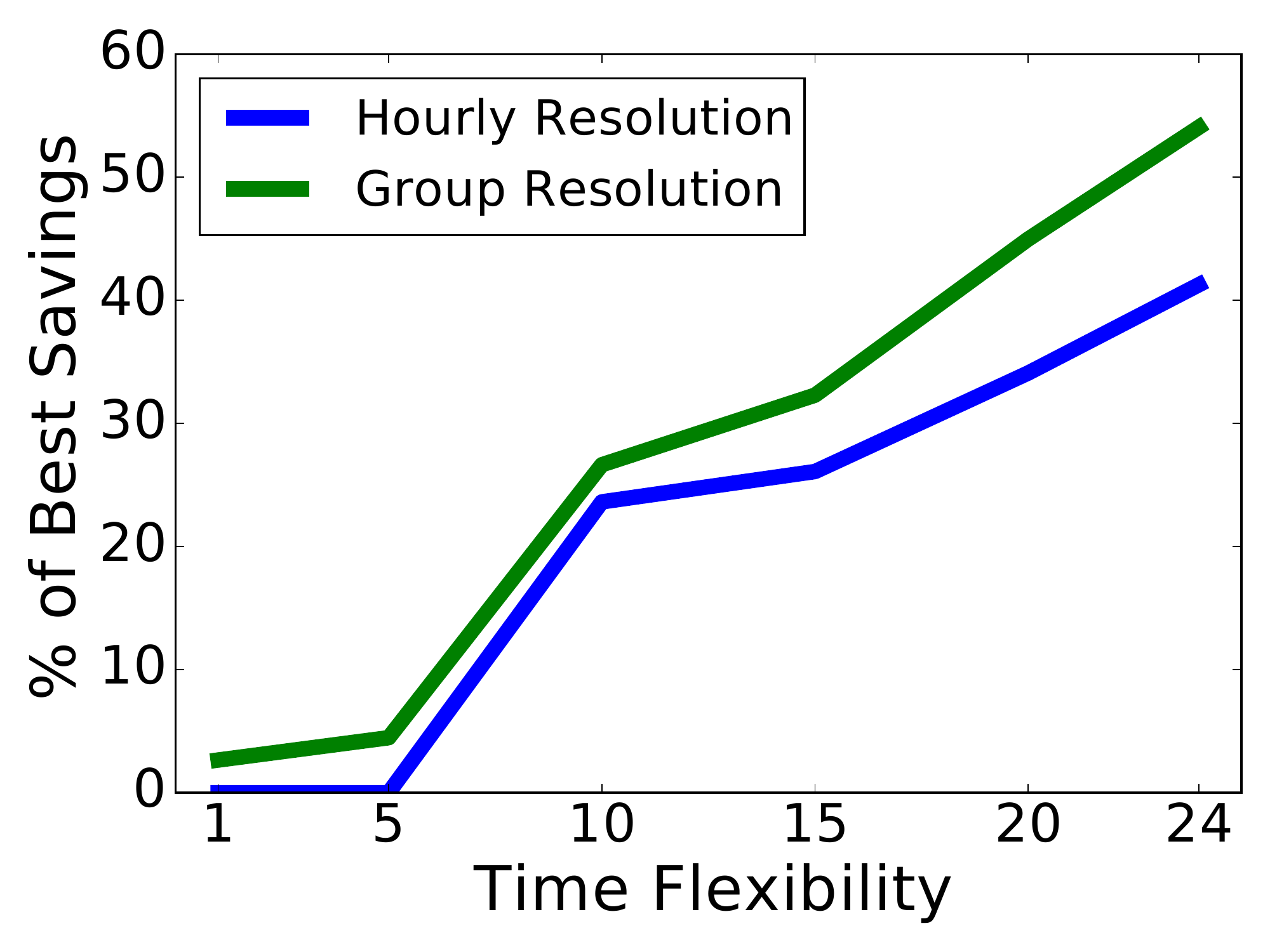}
	\caption{Best Saving: achieved for various time-flexibility}
	\label{fig1:LRSavingtime}
\end{figure}

Here, we quantify the utility of the device-level flexibility-based DR scheme. First, we find the theoretically maximum savings in regulation cost for a hypothetical 100\% accurate demand forecast model. Henceforth, we evaluate the percentage of the maximum savings that can be achieved by the proposed DR scheme at various data granularities and time flexibilities. Figure \ref{fig1:LRSavingtime} illustrates the best percentage savings at hourly and group resolution for varying time flexibilities. We can see that for both models, the best savings grow with increasing time flexibility. The hourly and group resolution models achieve the best savings of 42\% and 54\% of the optimal saving for 24 hours time-flexibility, respectively. The experimental results demonstrate the financial viability of the DR scheme where market players can significantly reduce the volume of energy traded in the regulation market and the associated cost. However, the savings in regulation cost comes from the dynamic scheduling of predicted device-level flexible demand and the extent of the savings depends on the underlying forecast model. It is interesting to analyze the achievable device-level forecast accuracy and its relation to the savings. Hence, in the next section, we will evaluate the proposed LR model for device-level demand prediction and thereafter evaluate the correlation between the forecast accuracy and financial savings.

\subsection{Device-level Forecast}

\begin{figure}
	\includegraphics[width=.9\columnwidth, trim=0px 0px 0px 0px, clip=true]{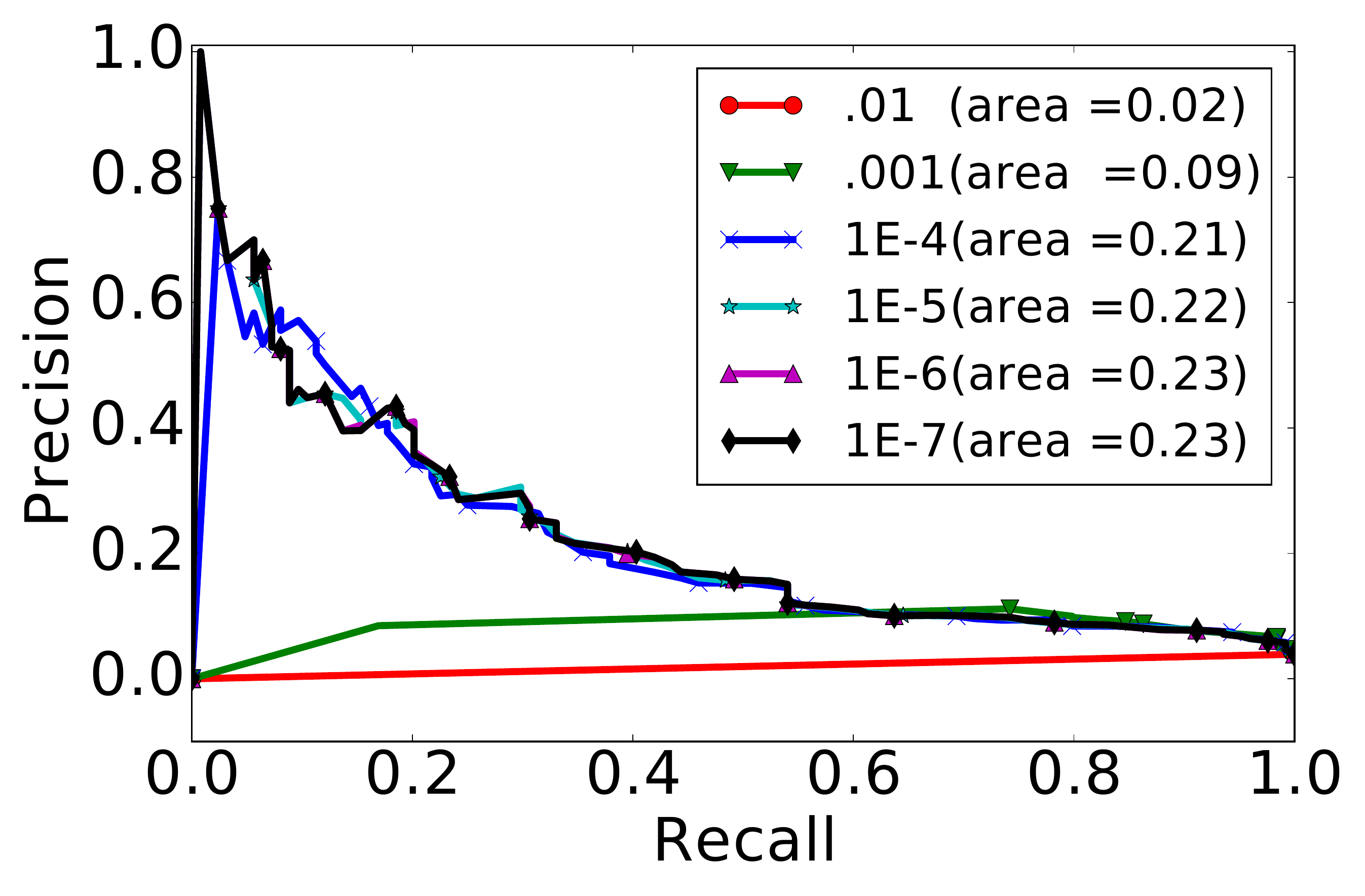}
	\caption{PR curve - for various $\lambda$ values (hourly).}
	\label{fig1:prLRWReg}
\end{figure}

In this section, we will analyze the performance of the classifier using the demand timeseries data for \textit{washer dryers}. First, we analyze the performance of the classifier for the hourly data granularity. 
Figure  \ref{fig1:prLRWReg} shows the  variation in the performance of forecast model when changing the regularization value $\lambda$.
The performance of the model degrades with increasing $\lambda$ values, mainly because increasing the penalty drives parameters $\theta$ to zero and deselect most of the features in $x$ (feature vector). 
The best regularization parameter for the forecast model is estimated via Cross-validation over each $\lambda$ value.

\begin{figure}
	\centering
	\includegraphics[width=.9\columnwidth, trim=0px 0px 0px 0px, clip=true]{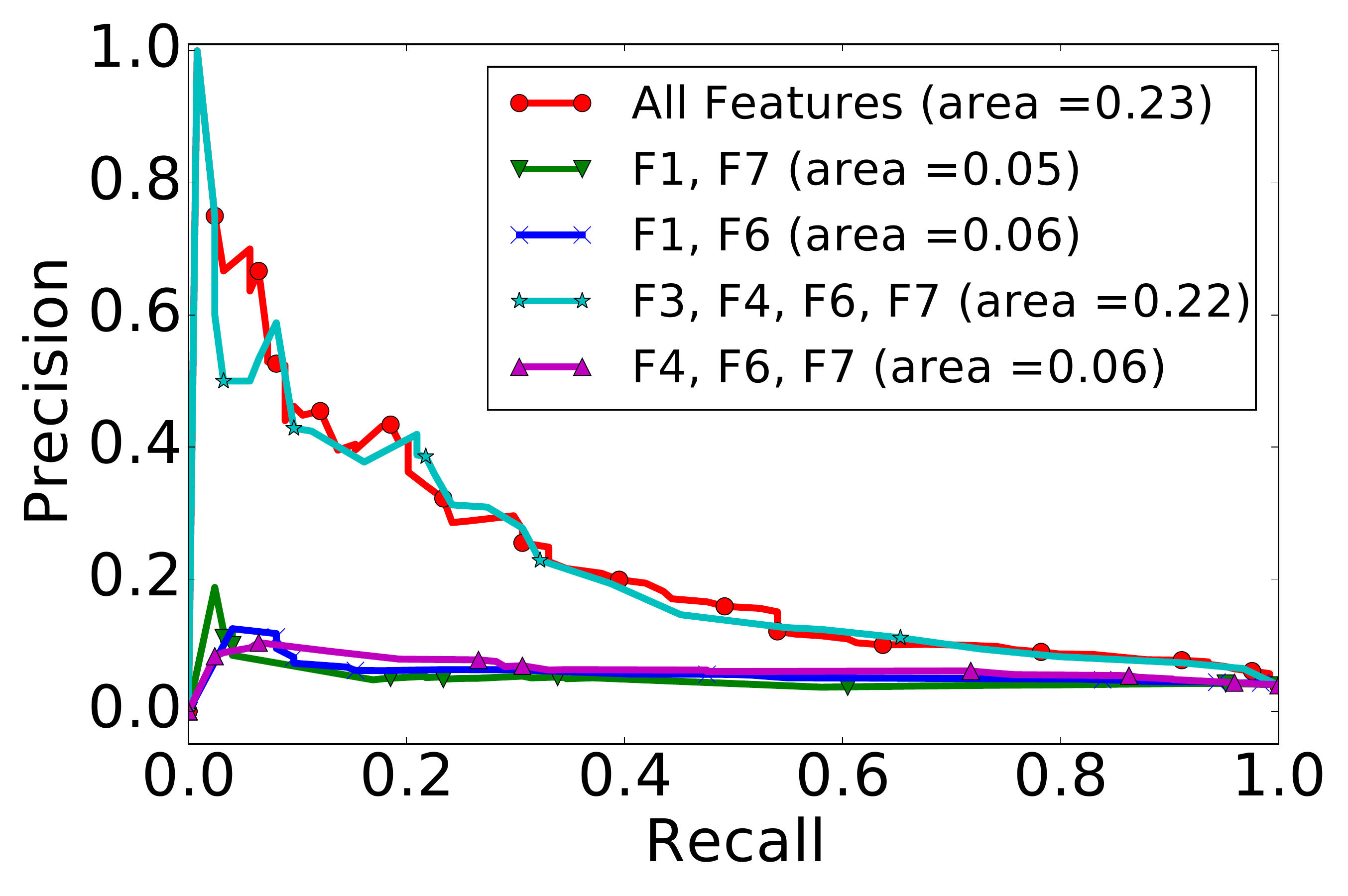}
	\caption{PR curve - for various feature sets (hourly).}
	\label{fig1:prLRWFeat}
\end{figure}
\begin{figure}
	\begin{minipage}{0.235\textwidth}
		\includegraphics[width=1\columnwidth, trim=0px 0px 0px 0px, clip=true]{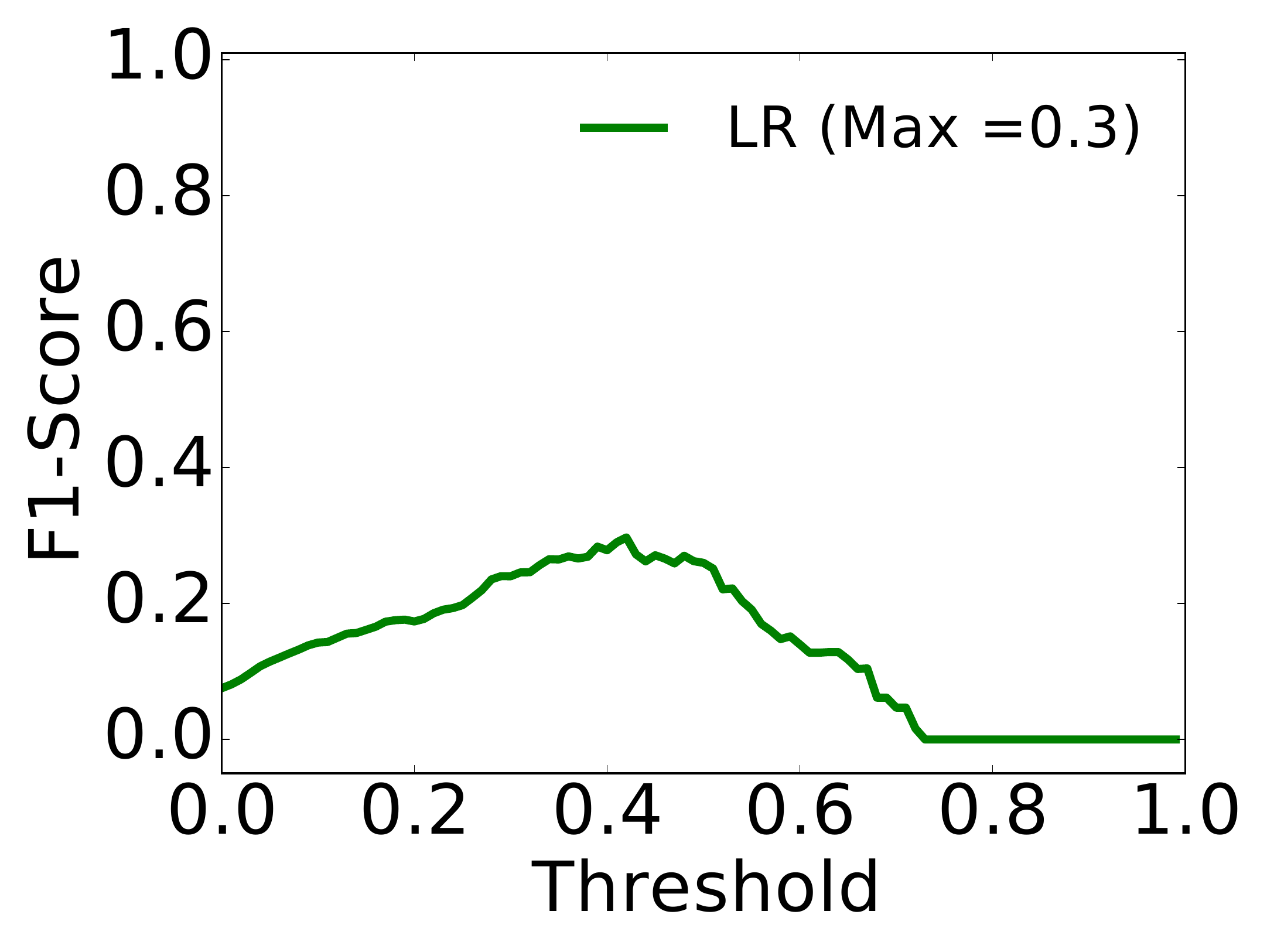}
		\caption{F1-Score for the classifier (hourly).}
		\label{fig1:F1All}
	\end{minipage}\hfill%
	\begin{minipage}{0.235\textwidth}
		\centering
		\includegraphics[width=1\columnwidth, trim=10px 10px 10px 10px, clip=true]{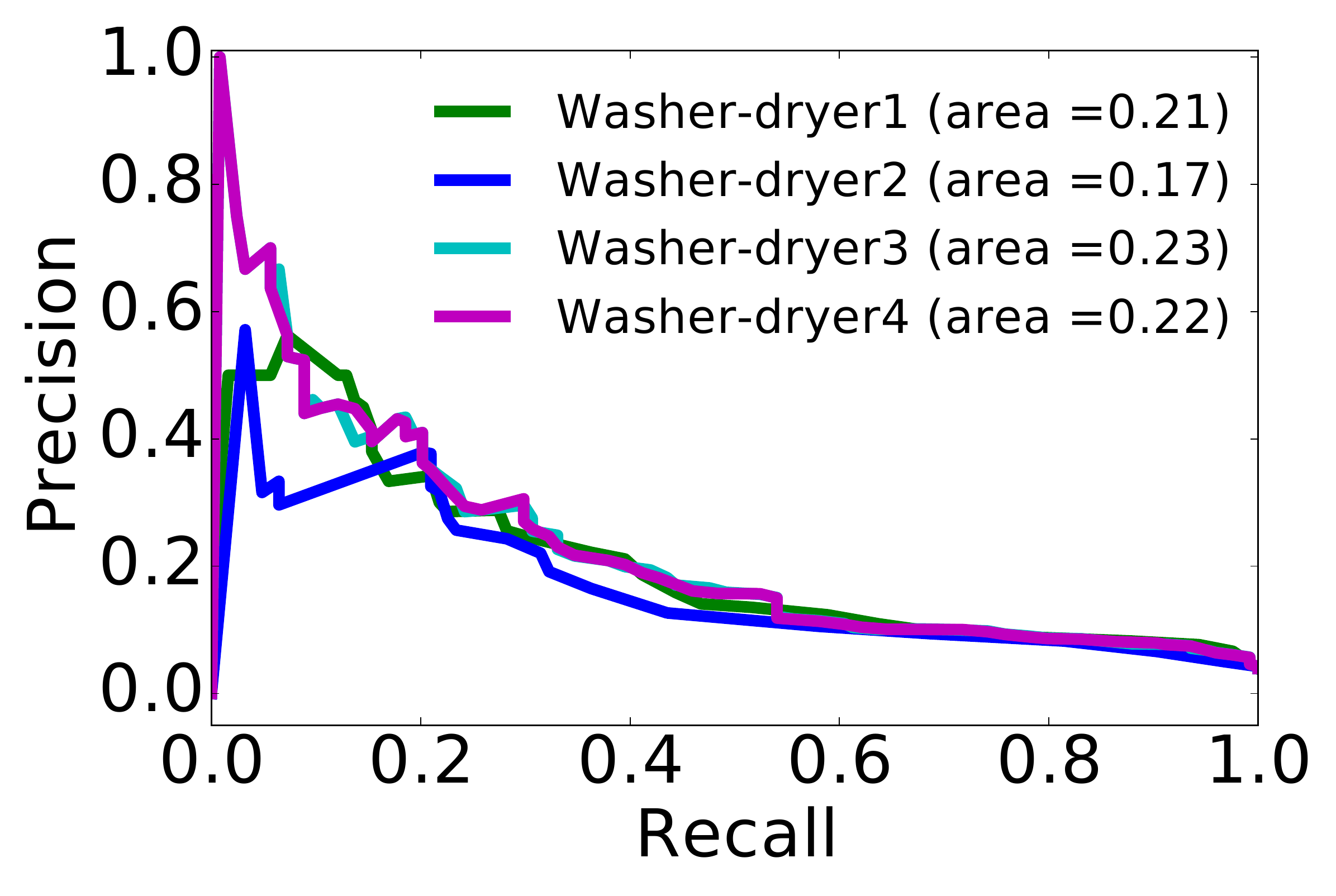} 
		\caption{PR curve - across washer dryers.}
		\label{fig:washers}
	\end{minipage}\hfill%
\end{figure}

 
The cross-validation gives the best average performance with $\lambda = 1E-6$  with an AUC of 0.23.
Further, Figure \ref{fig1:prLRWFeat} shows the performance of the classifier for various sets of features $x^{\{m\}}$, and the model achieved the best performance with the complete set of all extracted features. Thus, we argue that the best strategy is to feed a classifier with all the features and tune the model correctly so that it self-selects the most relevant ones.
As shown in Figure \ref{fig1:prLRWReg}, fluctuating (nonlinear) PR-curve complicates determining the best probability threshold for the model. 
Thus, to select the threshold value that gives the best performance of the model, we analyze the F1-scores of the classifier as shown in Figure \ref{fig1:F1All}. The classifier achieves the best performance at a threshold value of 0.42. 

\begin{figure}
	\begin{minipage}{0.235\textwidth}
		\includegraphics[width=1\columnwidth, trim=10px 10px 10px 10px, clip=true]{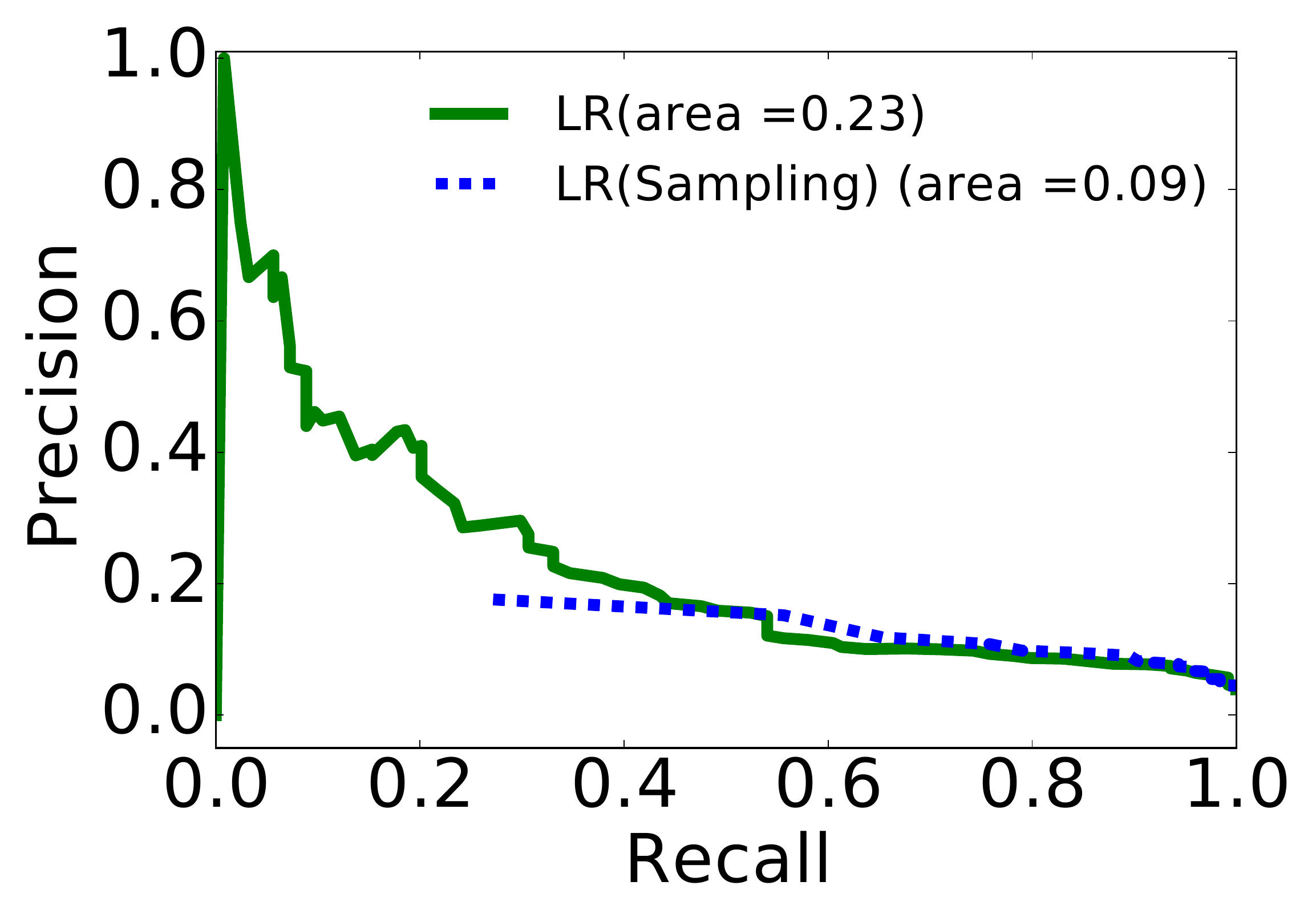}
		\caption{PR curve - for oversampling (hourly).}
		\label{fig1:PRLRWSAM}
	\end{minipage}\hfill%
	\begin{minipage}{0.235\textwidth}
		\centering
		\includegraphics[width=1.00\columnwidth, trim=10px 10px 10px 10px, clip=true]{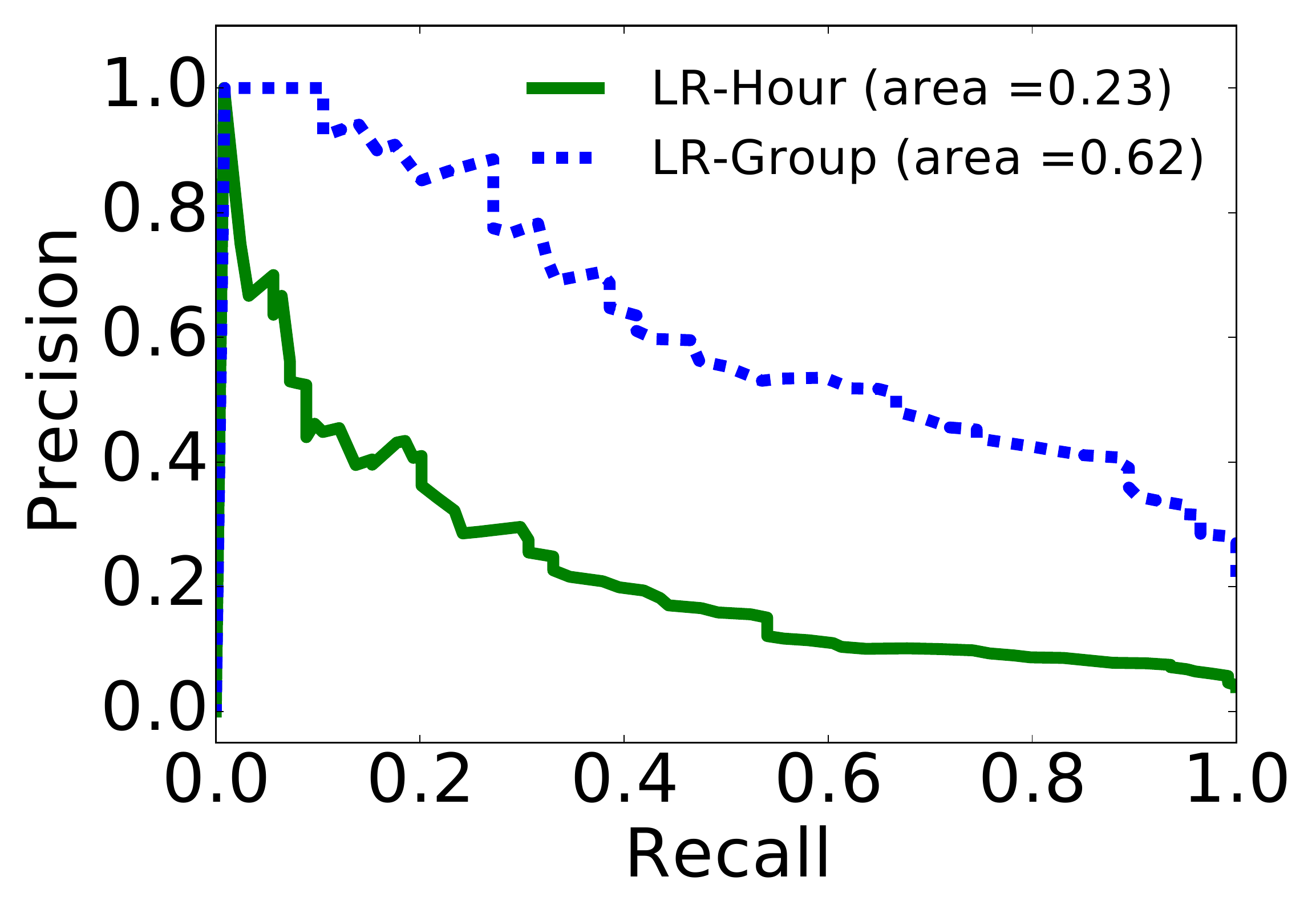}
		\caption{PR curve -(hourly vs group).}
		\label{fig1:PRLRWGRP}
	\end{minipage}\hfill%
\end{figure}

Further, as shown in Figure \ref{fig:washers}, the quite low AUC achieved by LR for \textit{washer dryers}, shows the stochasticity associated with device-level demands. 
Nevertheless, the comparable performance across the two devices illustrates that the proposed device-level forecast model is generalizable. 
The lower performance of the classifier is typically also due to a very high percentage of the majority class. Thus, we evaluate the classifier with oversampling of the minority class, shown in Figure \ref{fig1:PRLRWSAM}. 
However, oversampling increases the sensitivity of the classifier towards the minority class which further degrades the performance giving more FPs. 

\begin{figure}
	\begin{minipage}{0.225\textwidth}
		\includegraphics[width=1.00\columnwidth, trim=10px 10px 10px 10px, clip=true]{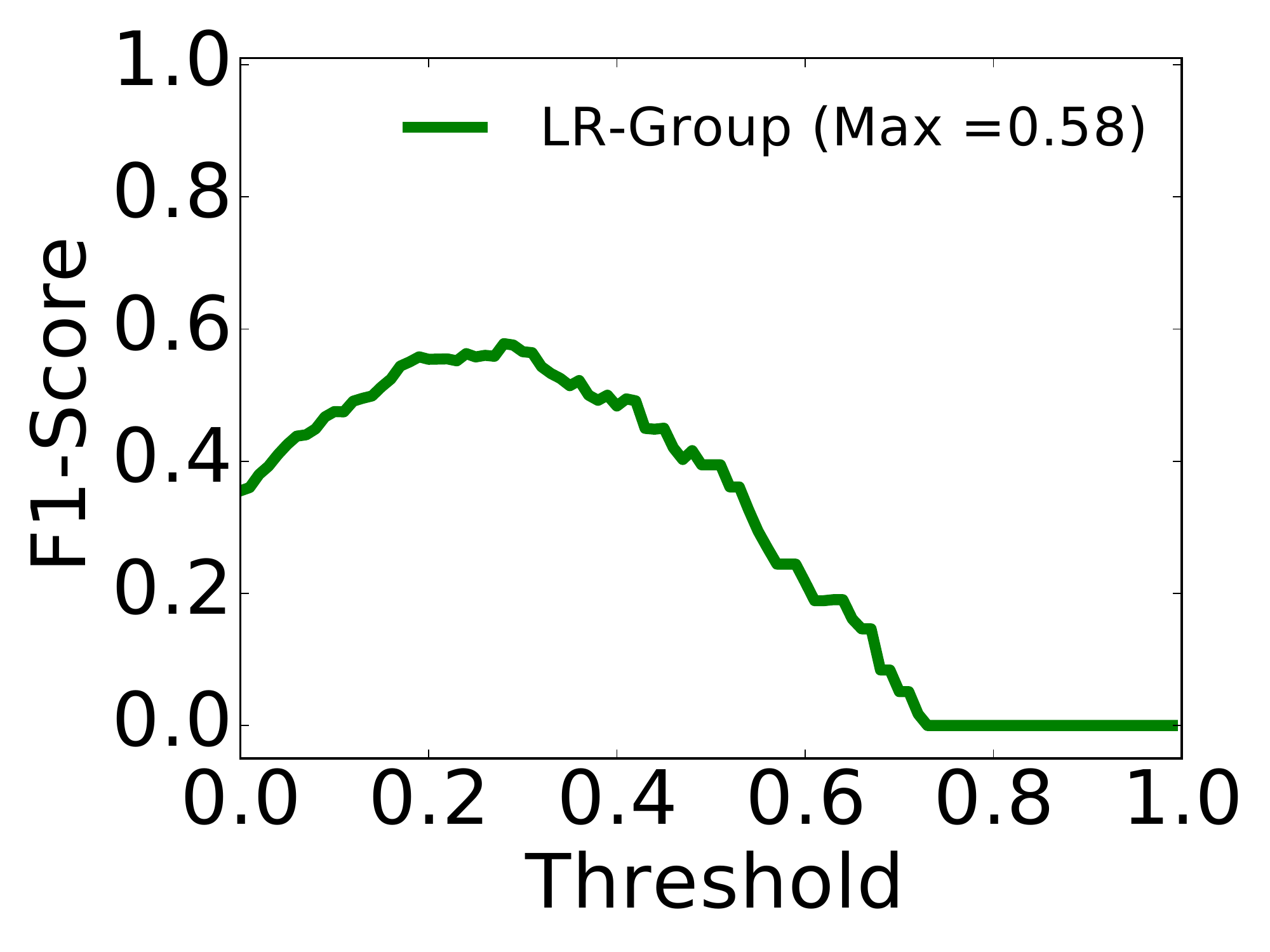}
		\caption{F1-Score for the classifier(group).}
		\label{fig1:F1ALLGRP}
	\end{minipage}\hfill%
	\begin{minipage}{0.245\textwidth}
		\centering
		\includegraphics[width=1\columnwidth, trim=10px 10px 10px 10px, clip=true]{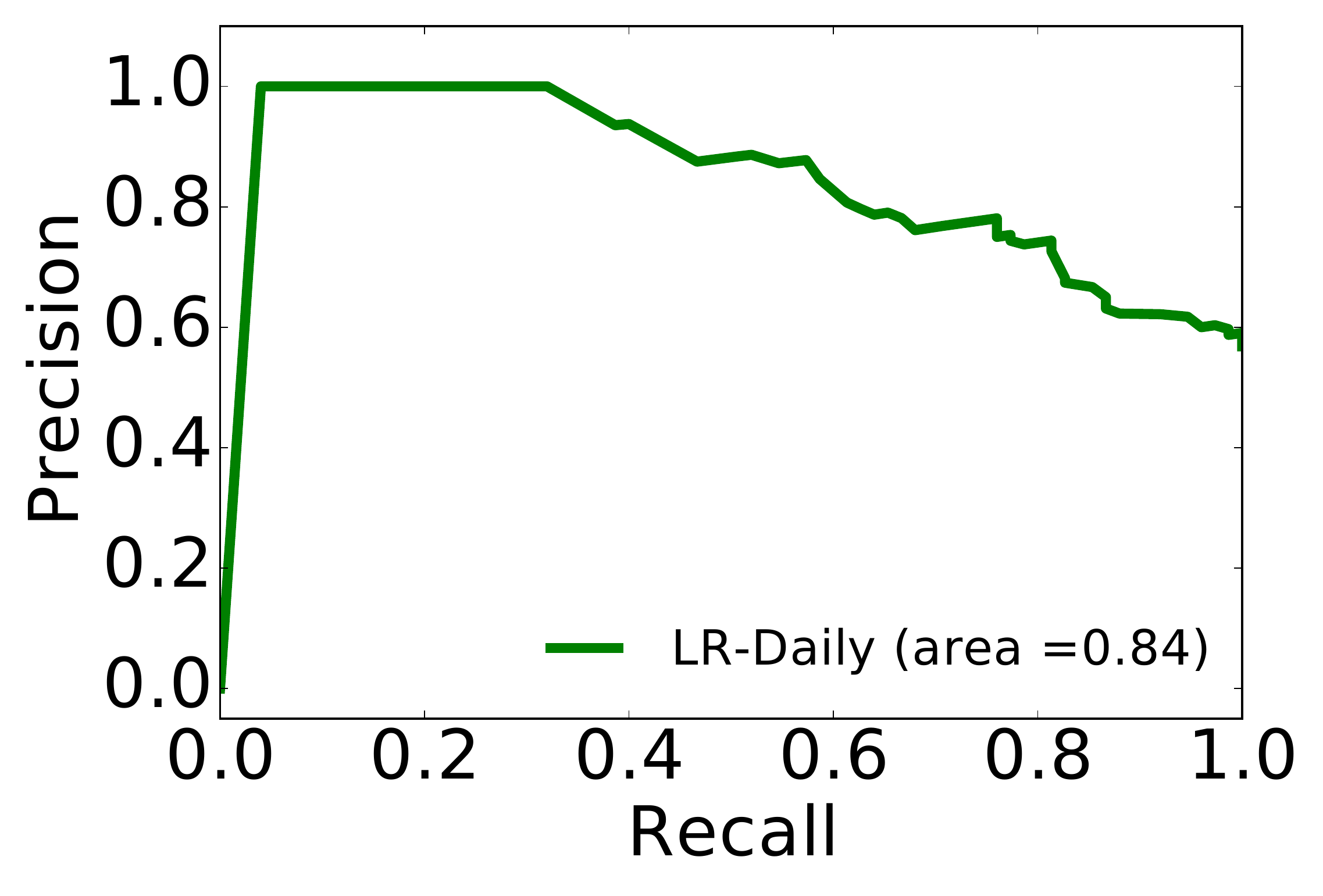}
		\caption{Performance of classifier (daily).}
		\label{fig1:PRALLDAILY}
	\end{minipage}\hfill%
\end{figure}

Figures \ref{fig1:PRLRWGRP} compares the performance of the classifier on hourly and group resolution, and Figure \ref{fig1:F1ALLGRP} shows the best F1-score achieved by the classifier at the group resolution.
The figure shows that the classifier has a significantly better performance at the group resolution with an improvement in AUC of 0.39. 
This result suggests that device usage patterns are more repetitive in a cluster of hours, e.g., the user frequently activates a \textit{washer dryer} in group $g_3$ (4 PM-12 AM) depending on his/her presence at home. 
Figure \ref{fig1:PRALLDAILY} shows the performance of the classifier at a daily resolution. The figure clearly demonstrate that the classifier achieve the best performance at a daily resolution with an AUC of 0.84. Moreover, we can conclude that, at a device-level, the predictability increases with the data aggregation level.

The above results exhibit the stochasticity associated with a device-level demand where it is hard to capture any patterns at an hourly resolution. 
In the absence of context information, the unusual behaviors in the usage patterns are wrongly represented by a forecasting model which degrades its overall performance. The results show that at the device-level, the forecast model achieves the best accuracy for daily resolution at the cost of a significant loss of demand flexibility, whereas, group resolution is found to provide a good trade-off between forecast accuracy and available demand flexibility. 
However, we would like to emphasize that our primary goal is not to design a highly accurate forecast model, but is to evaluate the viability of flexibility market utilizing a device-level demand forecast in a stochastic environment. Therefore, in the next section, we will analyze the financial implication on the regulation market relative to the performance level of our forecast models.

\subsection{Savings in relation to Forecast Accuracy}
\begin{figure}{r}
	\centering
	\includegraphics[width=.95\columnwidth, trim=10px 10px 10px 10px, clip=true]{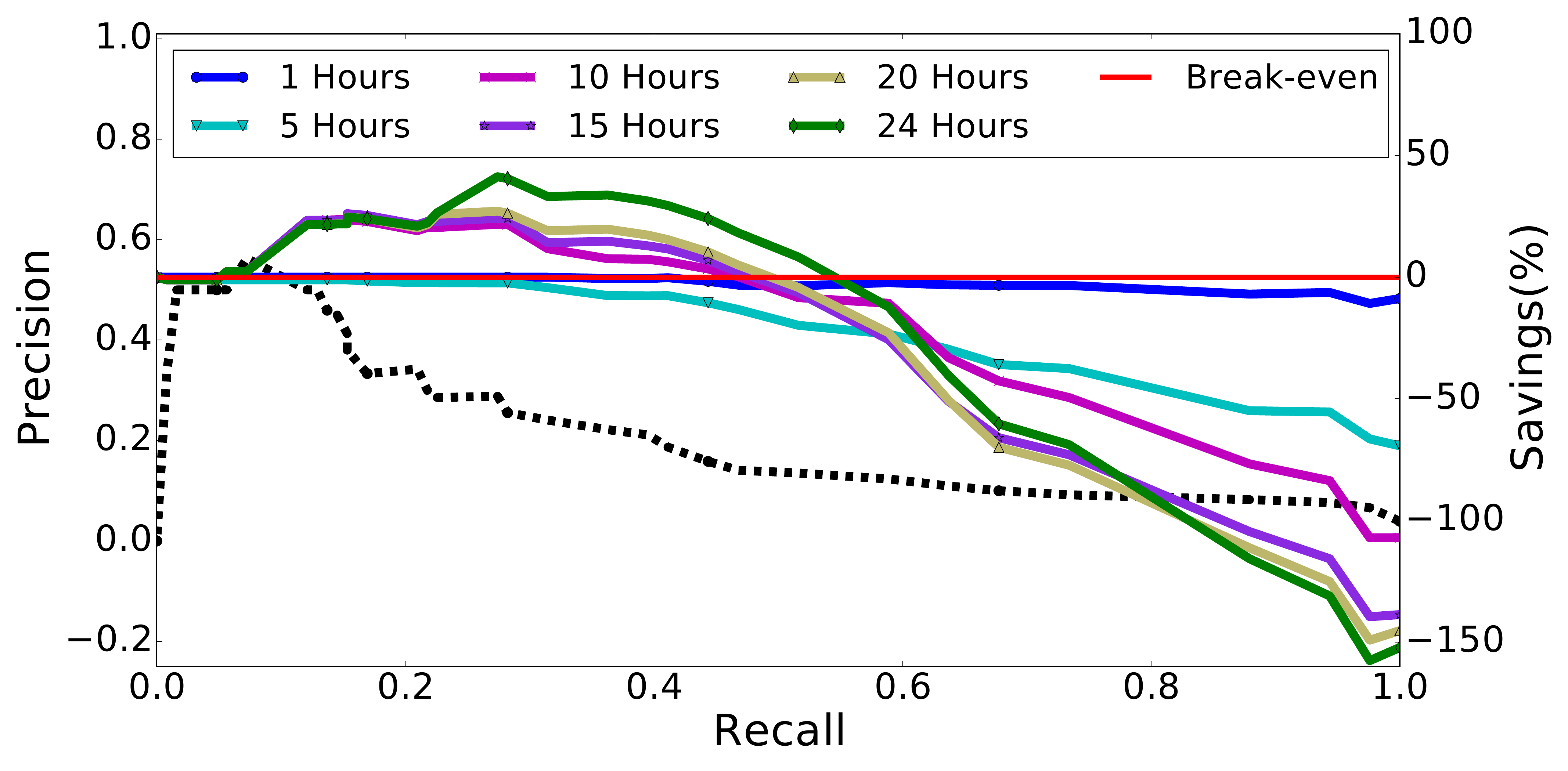}
	\caption{Savings from varying time flexibility (hourly).}
	\label{fig1:LRWSavingtime_hrl}
\end{figure} 

\begin{figure}{r}
	\centering
	\includegraphics[width=.95\columnwidth, trim=10px 10px 10px 10px, clip=true]{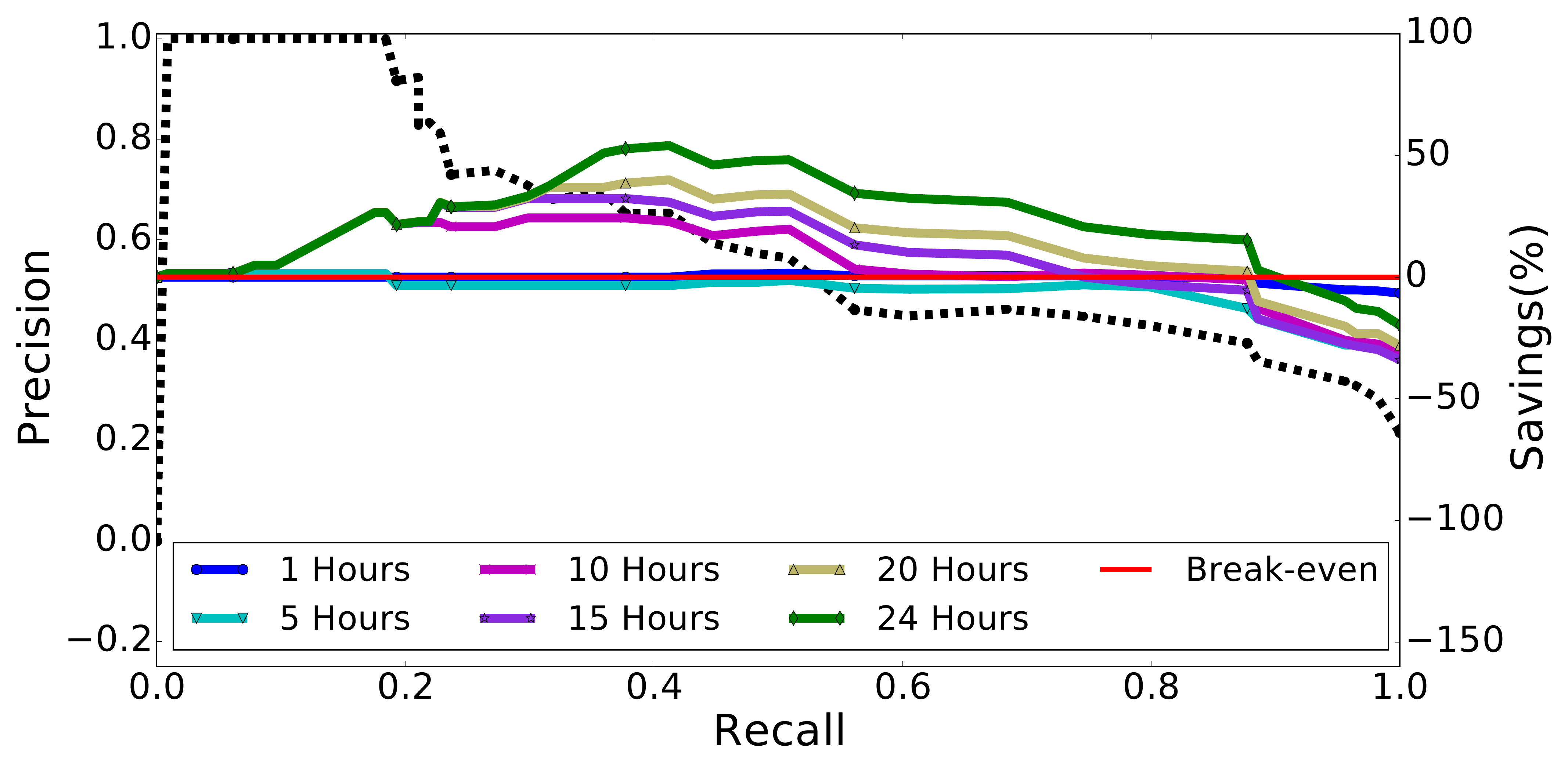}
	\caption{Savings from varying time flexibility (group).}
	\label{fig1:LRWSavingtime_grp}
\end{figure} 
Here, we will quantify the financial benefit (savings in regulation cost) of demand flexibility in relation to the achievable device-level demand forecast accuracy.  Figures \ref{fig1:LRWSavingtime_hrl} and \ref{fig1:LRWSavingtime_grp} show the breakdown of the savings in terms of precision and recall values for various data granularity and time flexibility.
In general, the savings is positive for a higher precision value, but the corresponding recall value determines the size, e.g., in Figure \ref{fig1:LRWSavingtime_hrl} the savings with a precision of 0.56 is 39\% less than with 0.30. 
The savings decreases due to the lower \textit{recall} where the loss due to the FP cases is higher than the gain due to the TP cases.
The figure illustrates that savings are positive only for a portion of the PR-curve and negative for the rest. The positive region represents precision-recall values with accuracy enough to cover the losses due to FN and FP cases.

\begin{figure}
	\begin{minipage}{0.235\textwidth}
		\centering
		\includegraphics[width=1\columnwidth, trim=10px 10px 10px 10px, clip=true]{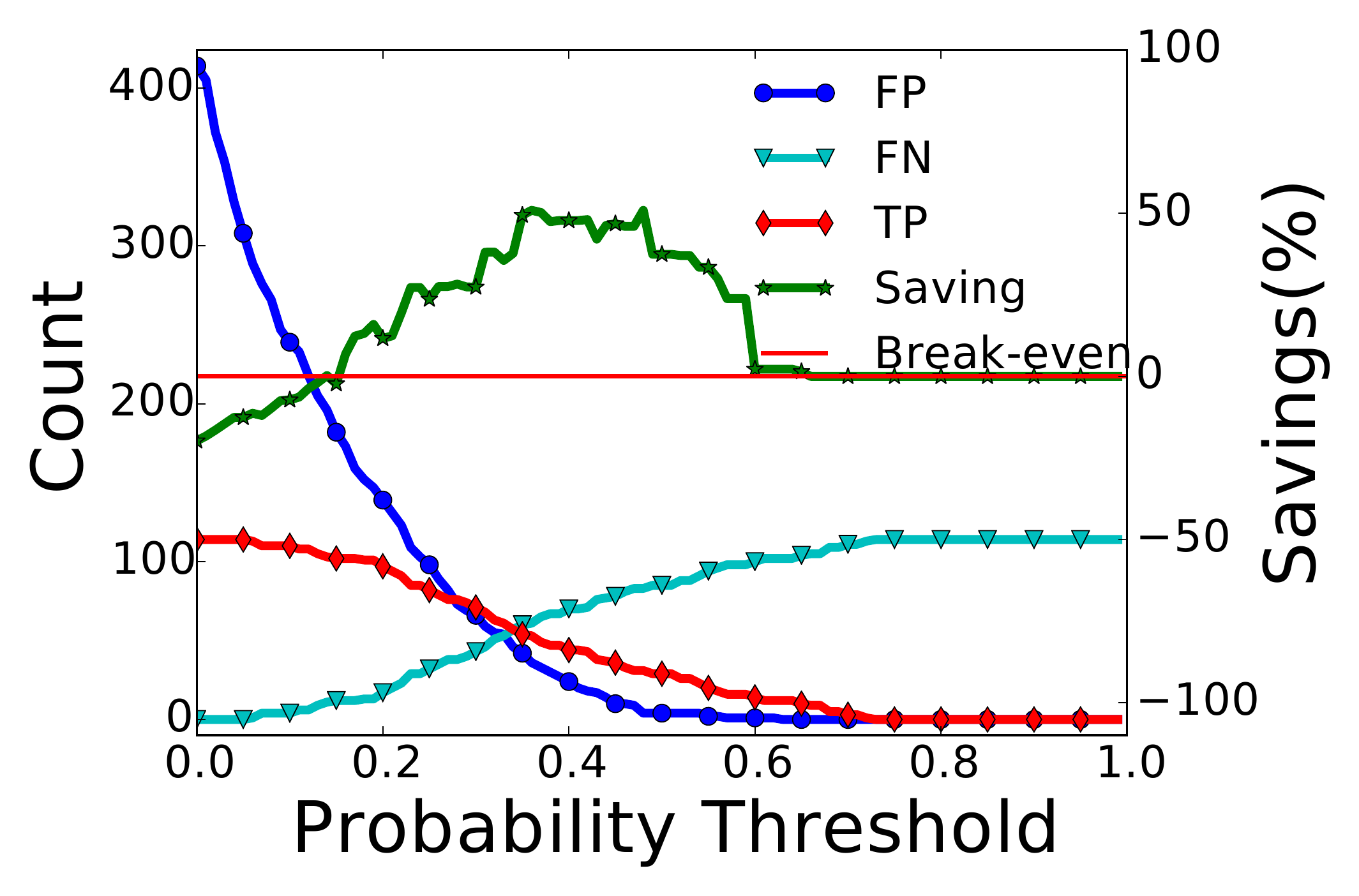}
		\caption{Saving relative to forecast category (group).}
		\label{fig1:FPTPLRWSaving}
	\end{minipage}\hfill%
	\begin{minipage}{0.235\textwidth}
		\centering
		\includegraphics[width=1\columnwidth, trim=10px 10px 10px 10px, clip=true]{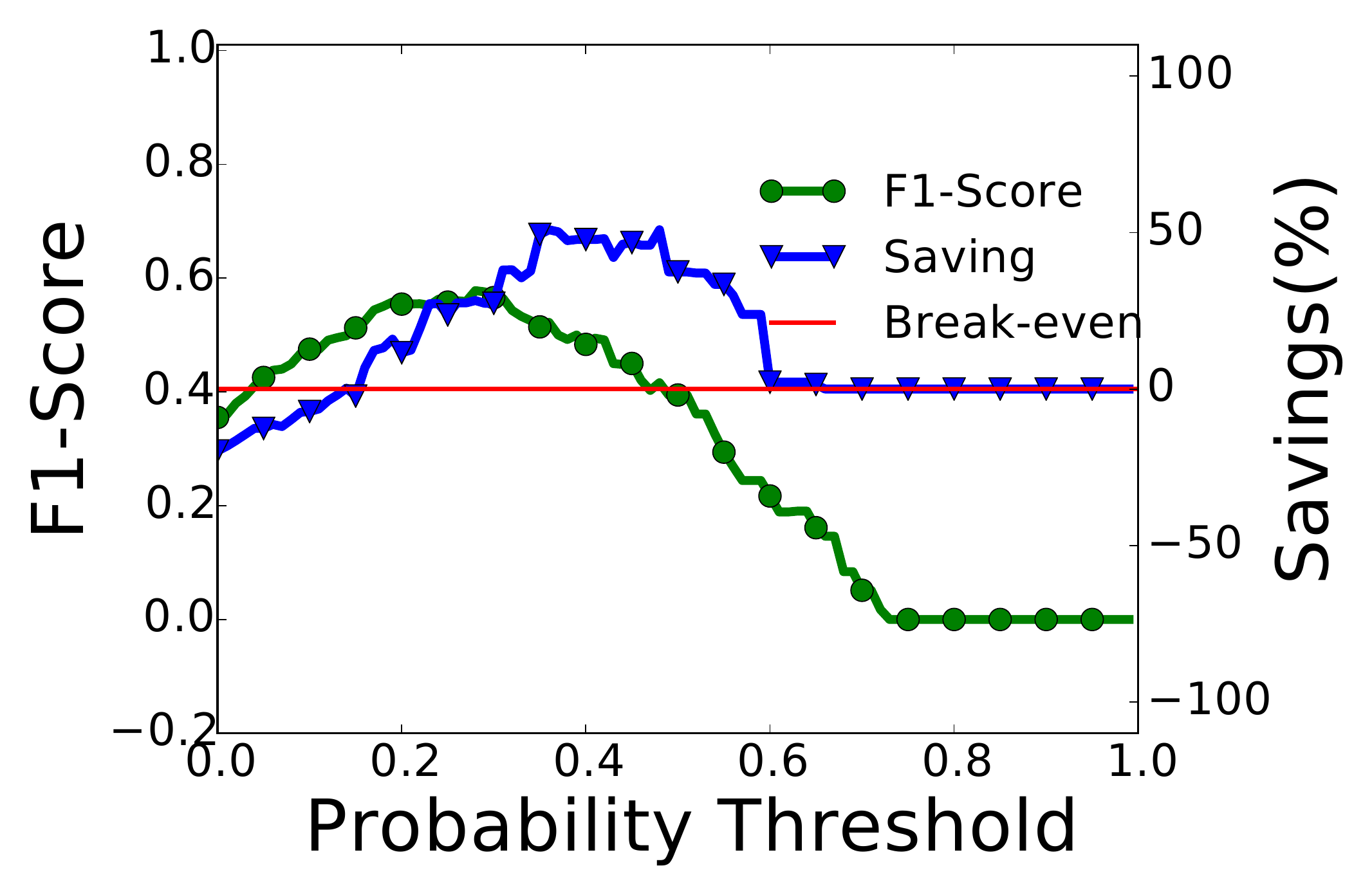}
		\caption{Saving relative to F1-score (group).}
		\label{fig1:F1LRWSaving}
	\end{minipage}%
\end{figure} 

Figures \ref{fig1:FPTPLRWSaving} shows the savings for various probability thresholds relative to the forecast categories, i.e., TP, FP, and FN. 
We can clearly see that at lower thresholds, the number of FPs is too high to generate any savings, e.g., at a probability threshold of 0.02 LR have FPs equivalent to 83\% of the total instances.
However, with an increase in the probability threshold, the FPs and TPs decreases as does the loss.  
The figures illustrate that a flexibility market can generate substantial savings even in the presence of a vast number of FPs. 
For example, a savings of 11\% for LR can be achieved even with 84\% of total forecasted flexible demand being false. These savings are mainly attributed to the change in regulation prices, where for some FPs the loss due to increase in up-regulation price at $i$ is lower than the gain due to a decrease in down-regulation price at $i+\tau$.

The savings do not follow the patterns of the PR-curve. 
This behavior creates difficulty in selecting a probability threshold value for a model that guarantees the positive savings, i.e., the probability threshold that gives the desired \textit{precision} and \textit{recall}. 
To this end, we evaluate the savings relative to the \textit{F1-score} at various probability threshold values, shown in Figure \ref{fig1:F1LRWSaving}.  
The figure shows that the savings follows the respective \textit{F1-score}, and the model achieves positive savings at a point of the highest \textit{F1-score}. 
Thus, the problem of setting the optimal probability threshold can be solved by selecting the value with the highest \textit{F1-score}. 
This rule of thumb is valid for all cases with significant savings, i.e., all experiments with best positive savings of $> 1.1\%$. 
\begin{figure}
	\begin{minipage}{0.235\textwidth}
		\centering
		\includegraphics[width=1\columnwidth, trim=10px 10px 10px 10px, clip=true]{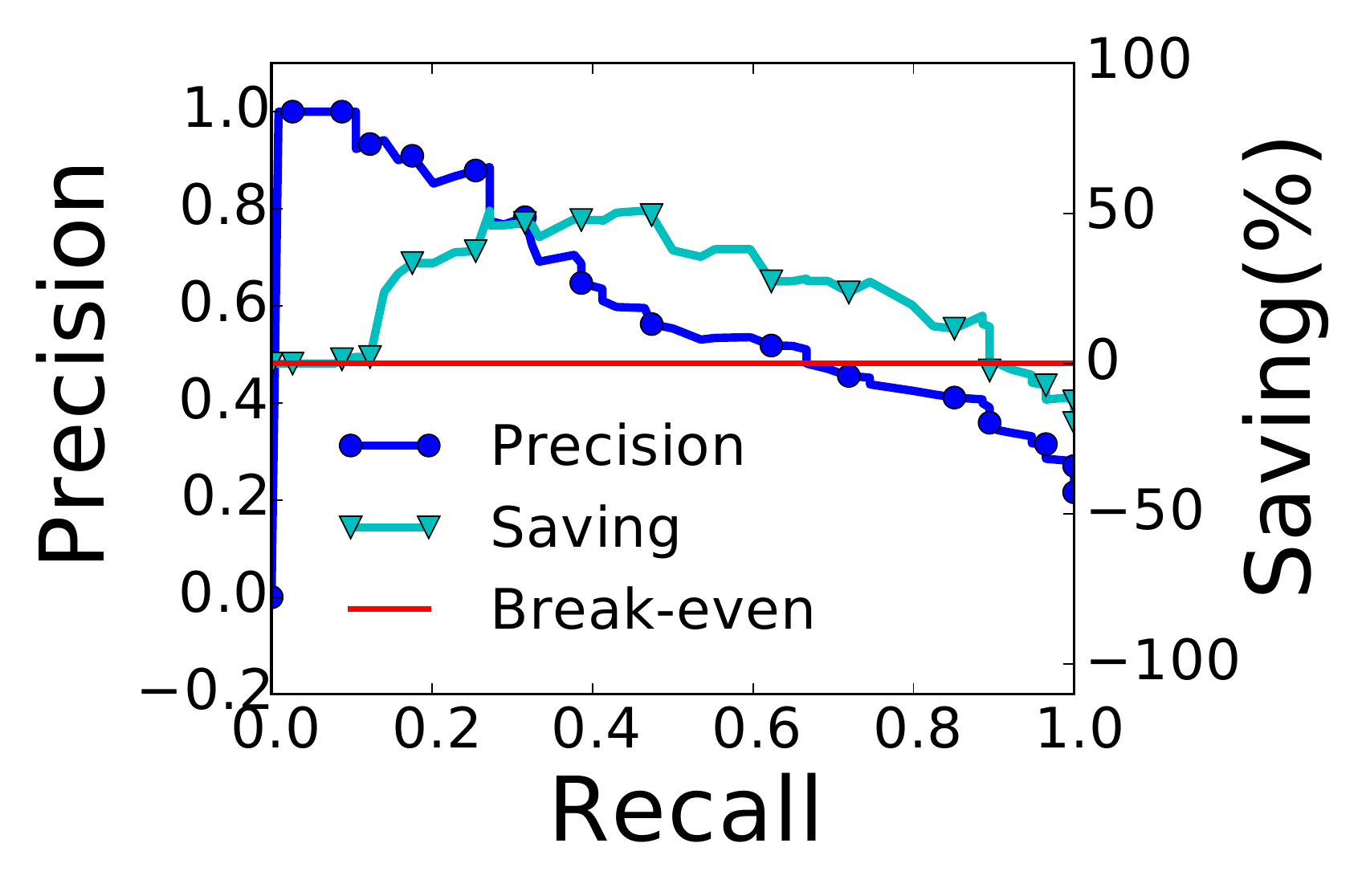}
		\caption{Saving relative to precision recall (group).}
		\label{fig1:prLRWSavingGroup}
	\end{minipage}\hfill%
	\begin{minipage}{0.235\textwidth}
		\centering
		\includegraphics[width=1\columnwidth, trim=10px 10px 10px 10px, clip=true]{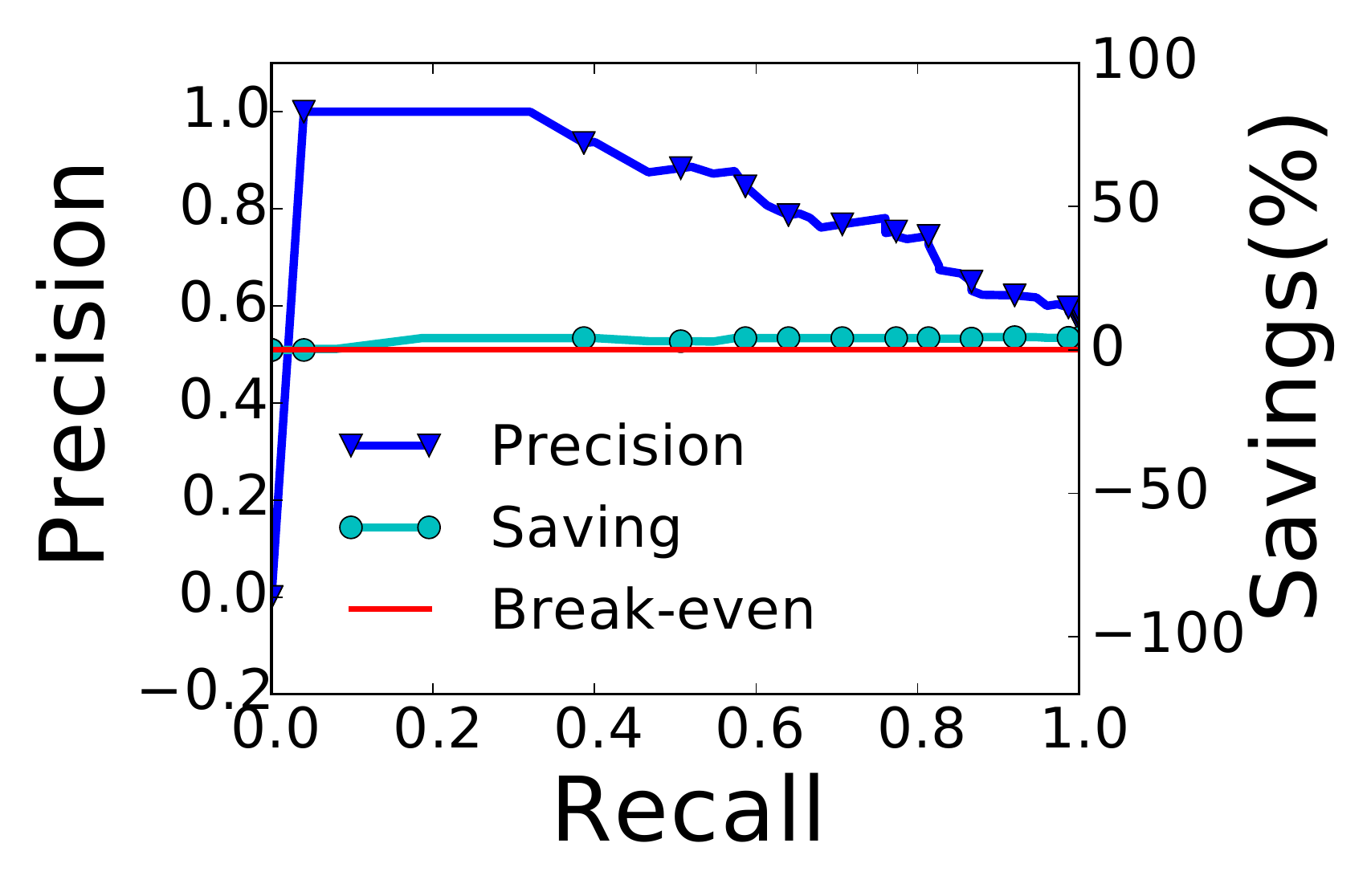}
		\caption{Saving relative to precision recall (daily).}
		\label{fig1:prLRWSavingDaily}
	\end{minipage}%
\end{figure} 

Figure \ref{fig1:prLRWSavingGroup} illustrates the savings from the demand flexibility utilizing the demand forecast at the group resolution. The figure demonstrates that the savings increase with the data aggregation, however, the increase is not proportional to the improvement in forecast accuracy.
For Example, we can see a 93\% improvement in the best \textit{F1-score}  for LR, but the savings increase only by 27\%. 
In addition, the figure demonstrates an extended positive region, which indicates that a market can generate savings for almost all values of the PR-curve. Further, a market (BRP) has a better confidence in the forecasted demand flexibility at a group resolution and is guaranteed to obtain savings from it.
The savings from the demand flexibility utilizing the demand forecast at the daily resolution is shown in Figure \ref{fig1:prLRWSavingDaily}. For the daily resolution, the savings is drastically reduced due to a decrease in the number of available flexible demands to be scheduled. Though, the savings is comparatively less, a market will never have a loss due to FP or FN flexible demands.

The above results show that the highest saving in regulation cost is achieved with a demand forecast model at the group resolution, nevertheless a market can also generate substantial savings at the hourly resolution. 
These results show that device-level \fbdr can be a promising tool to confront the challenges of integrating RES into the grid system. 
Moreover, an energy market can extract the benefit of device-level demand flexibility even in the presence of a large number of false predictions, i.e., FPs. However, the maximum proportion of FPs and FNs, i.e., the lower bound of precision and recall, that a market can sustain are specific to the market.

\section{Conclusion and Future Work}
In this paper, we analyzed the financial viability of flexibility-based DR at the atomic level, i.e., device-level, in relation to the achievable forecast accuracy.  In particular, we presented various features (attributes) to extract the device usage patterns. Thereupon, we assessed the feasibility of a widely used forecasting model, namely Logistic Regression for device level forecasting. Further, we formulated a set of equations to quantify the financial benefit and loss of demand flexibility, depending on the prediction categories represented by a contingency table, i.e., FN, FP, FN, and TP. Finally, we performed a number of experiments to evaluate the data aggregation level that provides the best financial reward to market players for adopting the proposed DR scheme.
The experimental results showed that, for the device-level demand forecast, financial gain for a market is much better than implied by the error metrics such as precision and recall. 
Market players can maximize their benefit of adopting flexibility-based DR scheme by utilizing a forecast model at the group resolution, where they can achieve the highest savings of 54\% of the optimal, 29\% higher than at hourly resolution. 
Further, the experiments showed that the savings in regulation cost grow with the increasing time flexibility. 
Furthermore, to set the probability threshold value that gives a near-optimal solution for a model, we presented a rule of thumb of selecting the threshold value with the highest F1-score.
Indeed, with a precision and recall of just 0.29 and 0.30, the market achieved regulation cost savings of 42\% of the theoretically optimal.

Important directions for future work are (1) experiments on a large number of households and devices, (2) evaluation of forecast models for a pool of similar devices, clustered based on various market criteria, and (3) evaluating the viability of flexibility-based DR in other energy markets.


\section*{Acknowledgments}
This work was supported in part by the TotalFlex project funded by the ForskEL program of Energinet.dk, the GoFLEX project funded under the Horizon 2020 programme, and the DiCyPS project funded by Innovation Fund Denmark.


\appendix
\section*{Appendix}
In this section, we will present the additional experimental results. We compare the performance of three different classifiers i) Logistic Regression with weighted class importance(LR) (used in the main section of the paper), ii)  Logistic Regression with Regularization (NLR), and iii) Pattern Matching (PM), for device level demand prediction and associated financial benefits. 
\section{Comparison of Model Performance}

\begin{figure}[!htb]
	\centering
	\begin{subfigure}{.25\textwidth}
		\centering
		\includegraphics[width=1\linewidth]{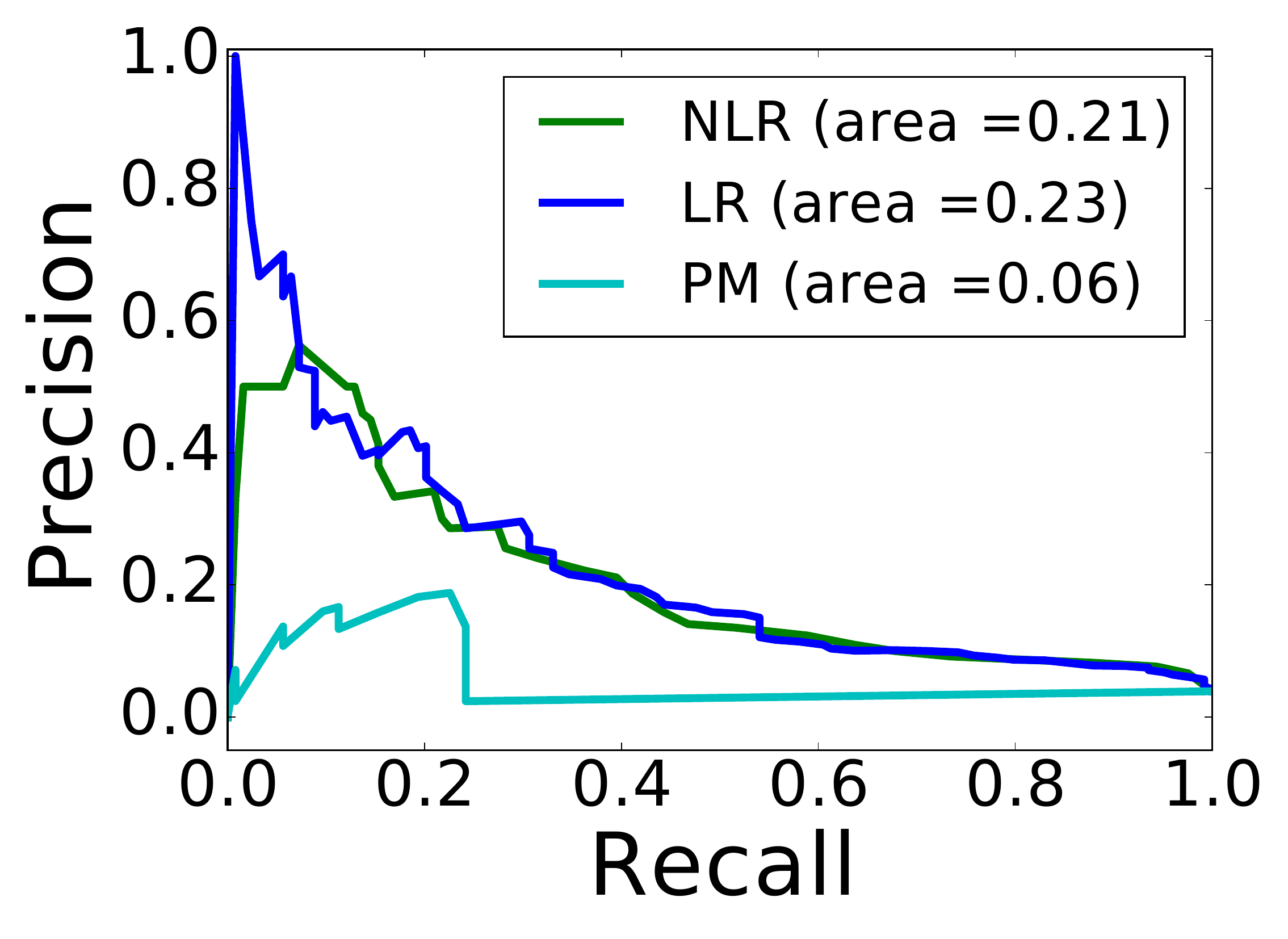}
		\caption{PR curve.}
		\label{fig1:PRAll1}
	\end{subfigure}%
	\begin{subfigure}{.25\textwidth}
		\centering
		\includegraphics[width=1\linewidth]{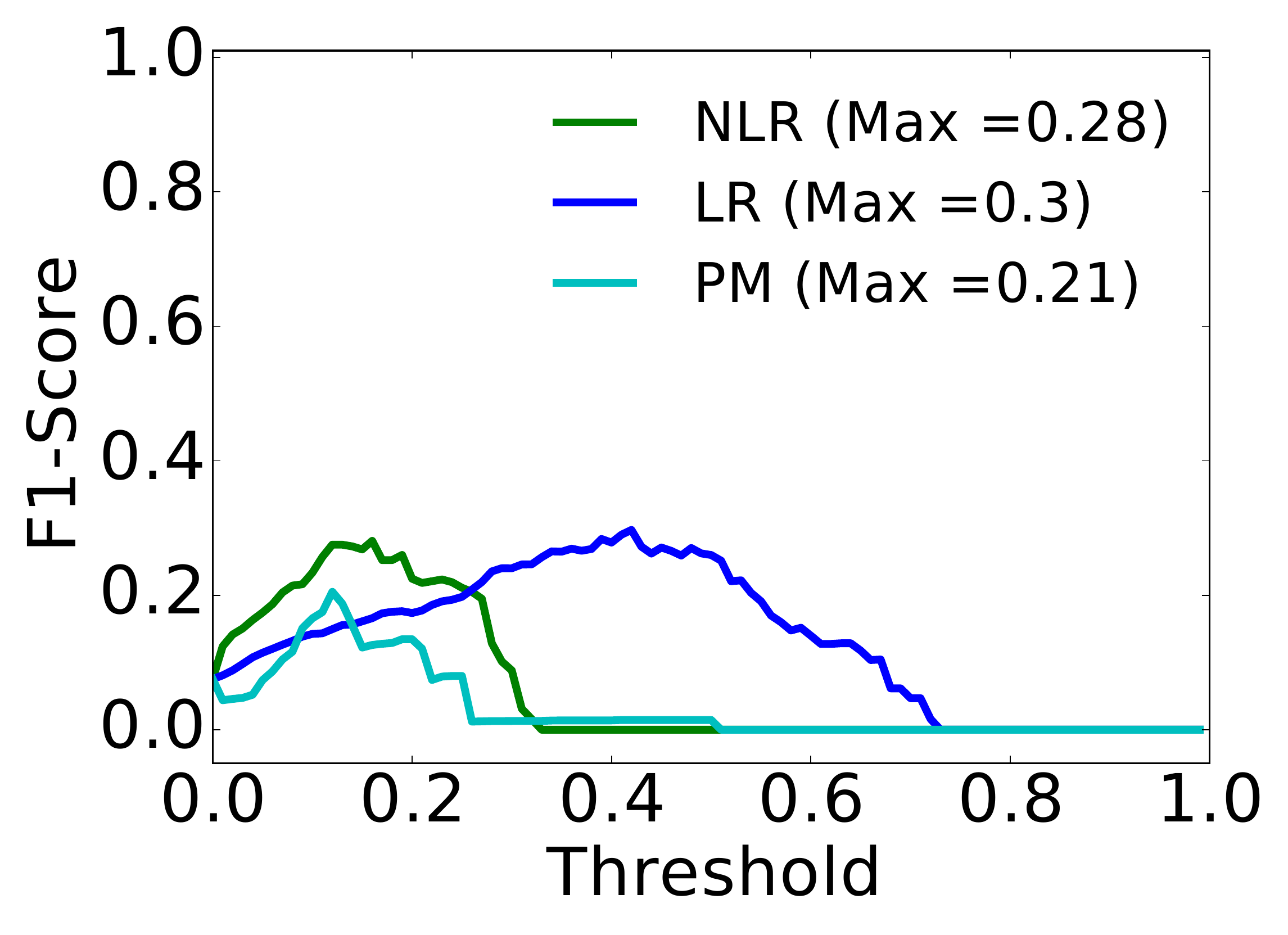}
		\caption{F1-Score}
		\label{fig1:F1All1}
	\end{subfigure}%
	\caption{Performance of classifiers (hourly).}
\end{figure}

The precision-recall curves for all 3 classifiers are shown in Figure \ref{fig1:PRAll1}. The figure shows that LR model has the highest area under the curve of 0.23 compared to 0.21 and 0.06 for NLR and PM, respectively. 
Further, the results show that the predicted class probabilities for NLR and PM are clustered in a small region, i.e., have lower prediction confidence.  
On the other hand, LR predicts the positive and negative class with higher confidence that gives a smooth precision-recall curve.  As discussed before, the lower confidence in prediction is due to the class imbalance and stochastic behaviors associated with the device-level demand. The lower prediction confidence creates fluctuating (non-linear) precision and recall curve as shown in Figure  \ref{fig1:PRAll1}. The Figure \ref{fig1:F1All1} illustrates that NLR, LR, and PM achieve the best performance at a threshold value of 0.16, 0.42, and 0.12, respectively. From the figures, we can see that none of the classifiers show a good performance at an hourly resolution, and a simple model such as PM has a performance comparable to a complex model (LR).

\begin{figure}[!htb]
	\centering
	\begin{subfigure}{.25\textwidth}
		\centering
		\includegraphics[width=1\linewidth]{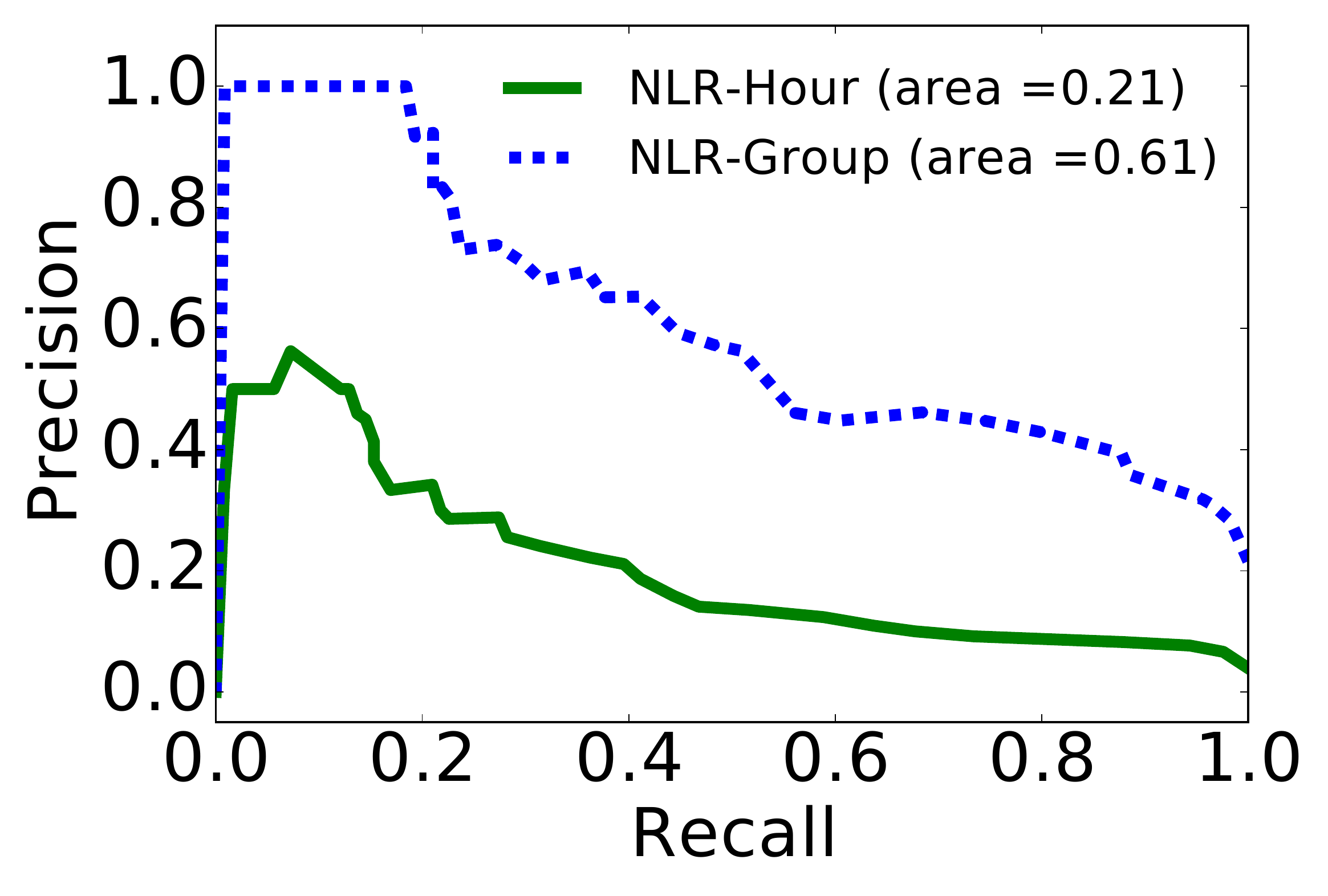}
		\caption{NLR}
		\label{fig1:PRLRGRP1}
	\end{subfigure}%
	\begin{subfigure}{.25\textwidth}
		\centering
		\includegraphics[width=1\linewidth]{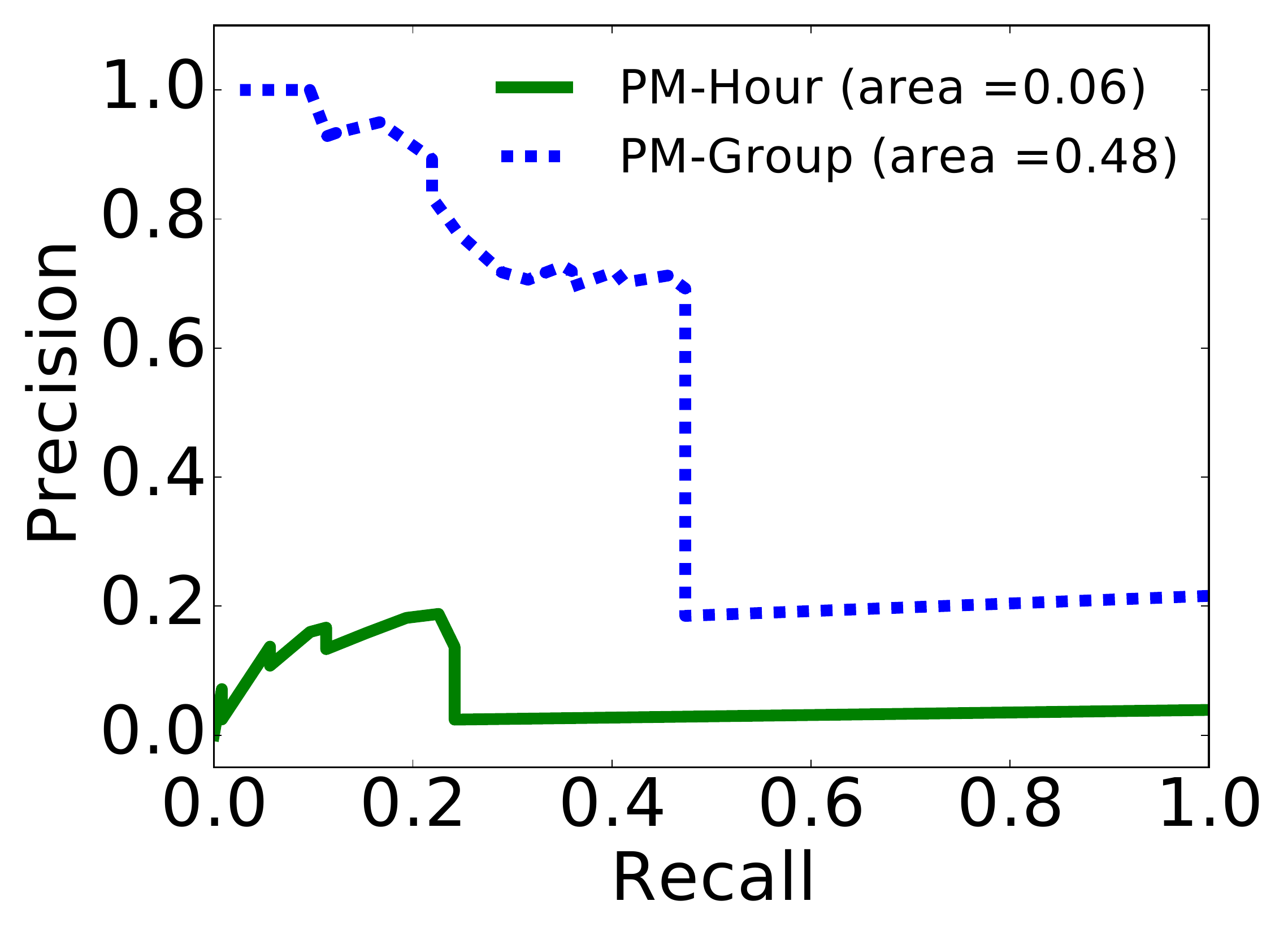}
		\caption{PM}
		\label{fig1:PRPMGRP1}
	\end{subfigure}%
	\caption{Performance of classifiers (hourly versus group).}
\end{figure} 

Figures \ref{fig1:PRLRGRP1}, \ref{fig1:PRLRWGRP}, and \ref{fig1:PRPMGRP1} compare the performance of the three classifiers on hourly and group resolution. As seen before, the NLR and PM also have a significantly better performance at the group resolution. Especially, PM reports the best performance improvement with an increase in AUC of 0.42.  Further, as with hourly resolution, the performances of NLR and LR are comparable with LR leading by only 2\%.

\begin{figure}[!htb]
	\centering
	\begin{minipage}{0.21\textwidth}
		\centering
		\includegraphics[width=1\linewidth]{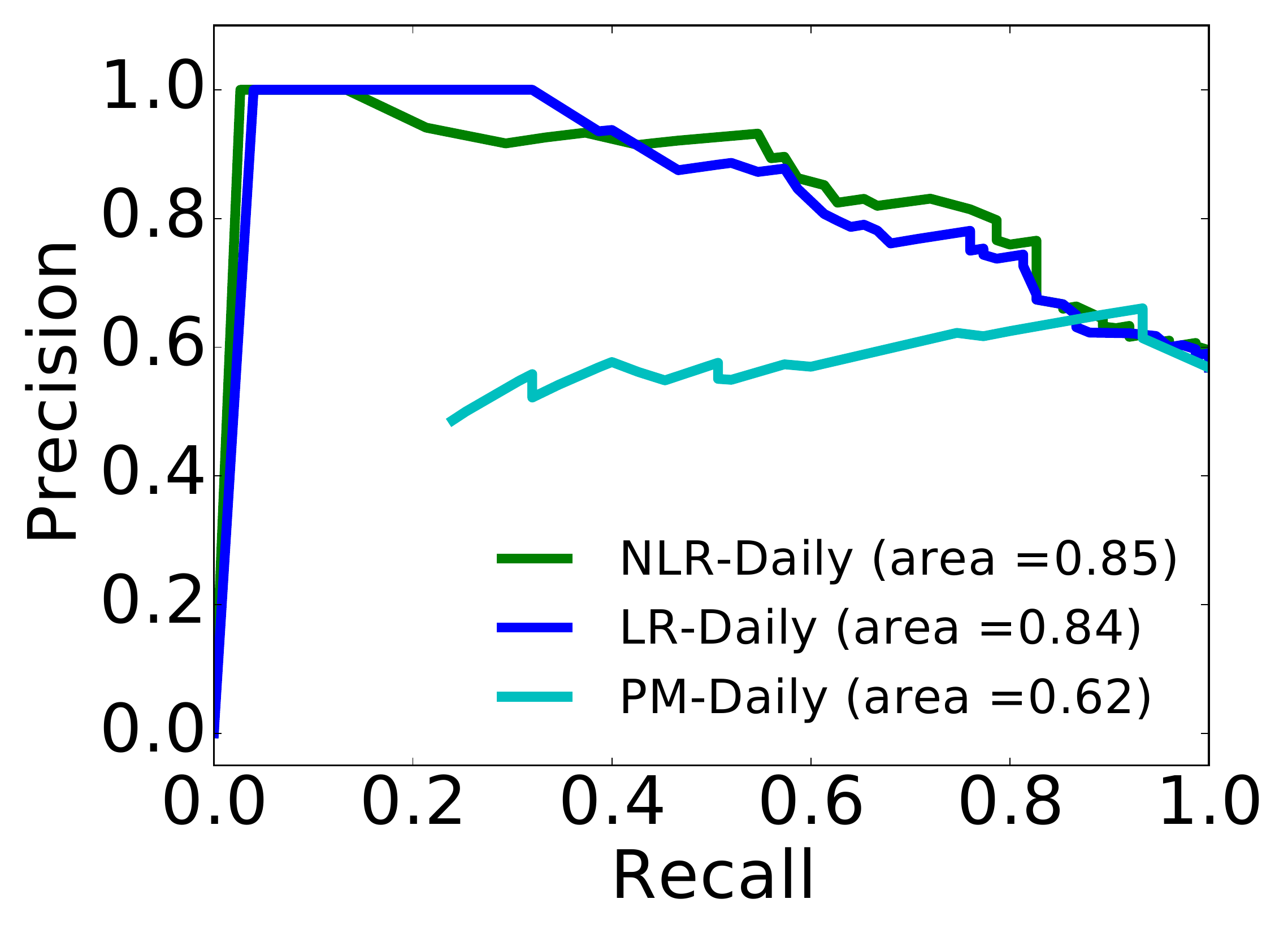}
		\caption{Performance of classifiers (daily).}
		\label{fig1:PRALLDAILY1}
	\end{minipage}\hfill%
	\begin{minipage}{0.26\textwidth}
		\centering
		\includegraphics[width=1\linewidth]{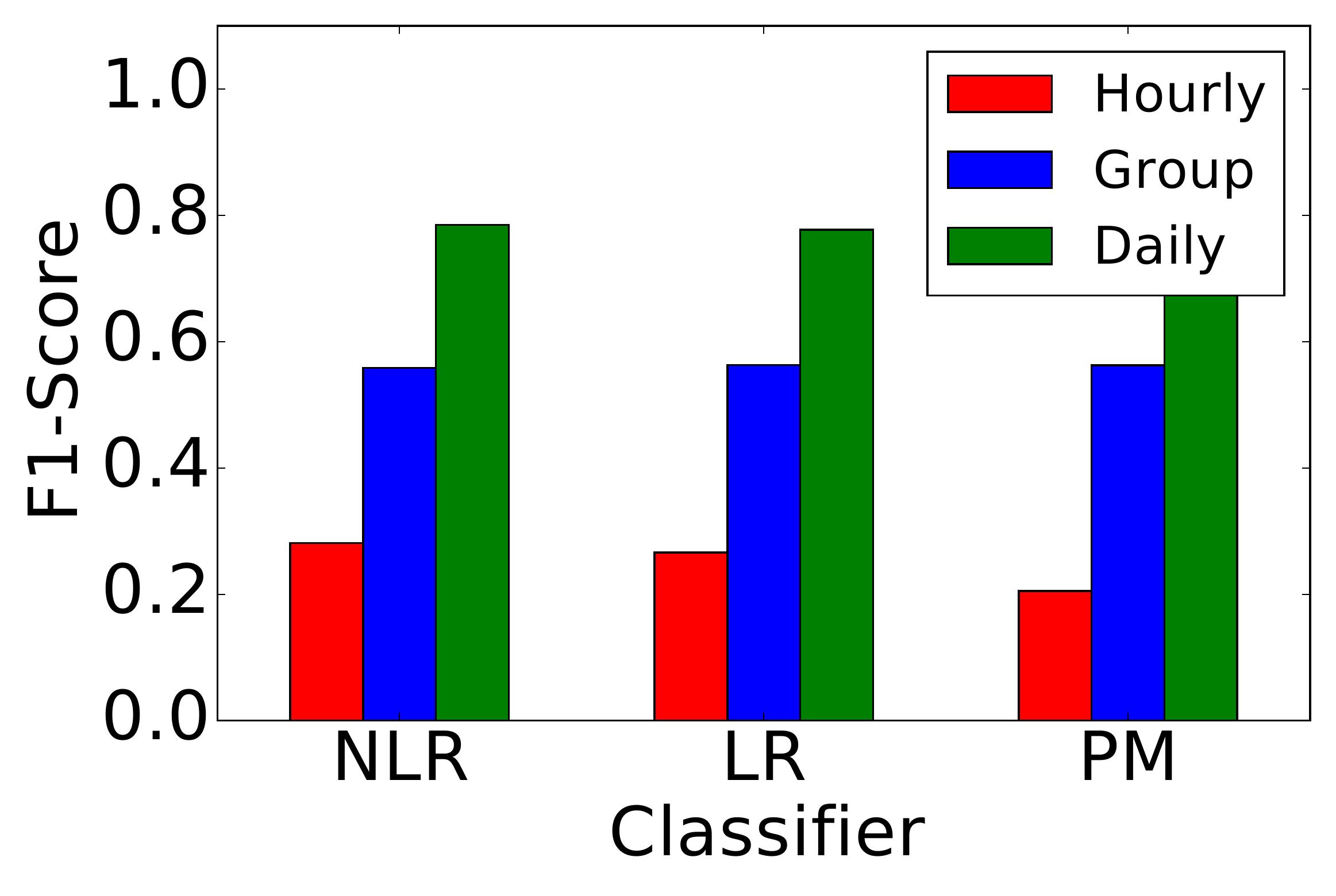}
		\caption{Comparison of classifiers - across all resolution.}
		\label{fig1:F1ALLDAILY1}
	\end{minipage}\hfill%
\end{figure} 

Figure \ref{fig1:PRALLDAILY1} compares the performance of the classifiers at a daily resolution, and Figure \ref{fig1:F1ALLDAILY1} compares the best F1-score.  
The figures demonstrate that the classifiers achieve the best performance at a daily resolution with AUC of 0.85, 0.84, 0.62 for NLR, LR, PM, respectively. At the daily resolution, the imbalance shift towards the positive class and the weighted measure does not contribute to the performance gain. 
Therefore, the performance of NLR surpasses LR.

\section{Comparison of Savings}
\begin{figure}[!htb]
	\centering
	\begin{subfigure}{.25\textwidth}
		\centering
		\includegraphics[width=1\linewidth]{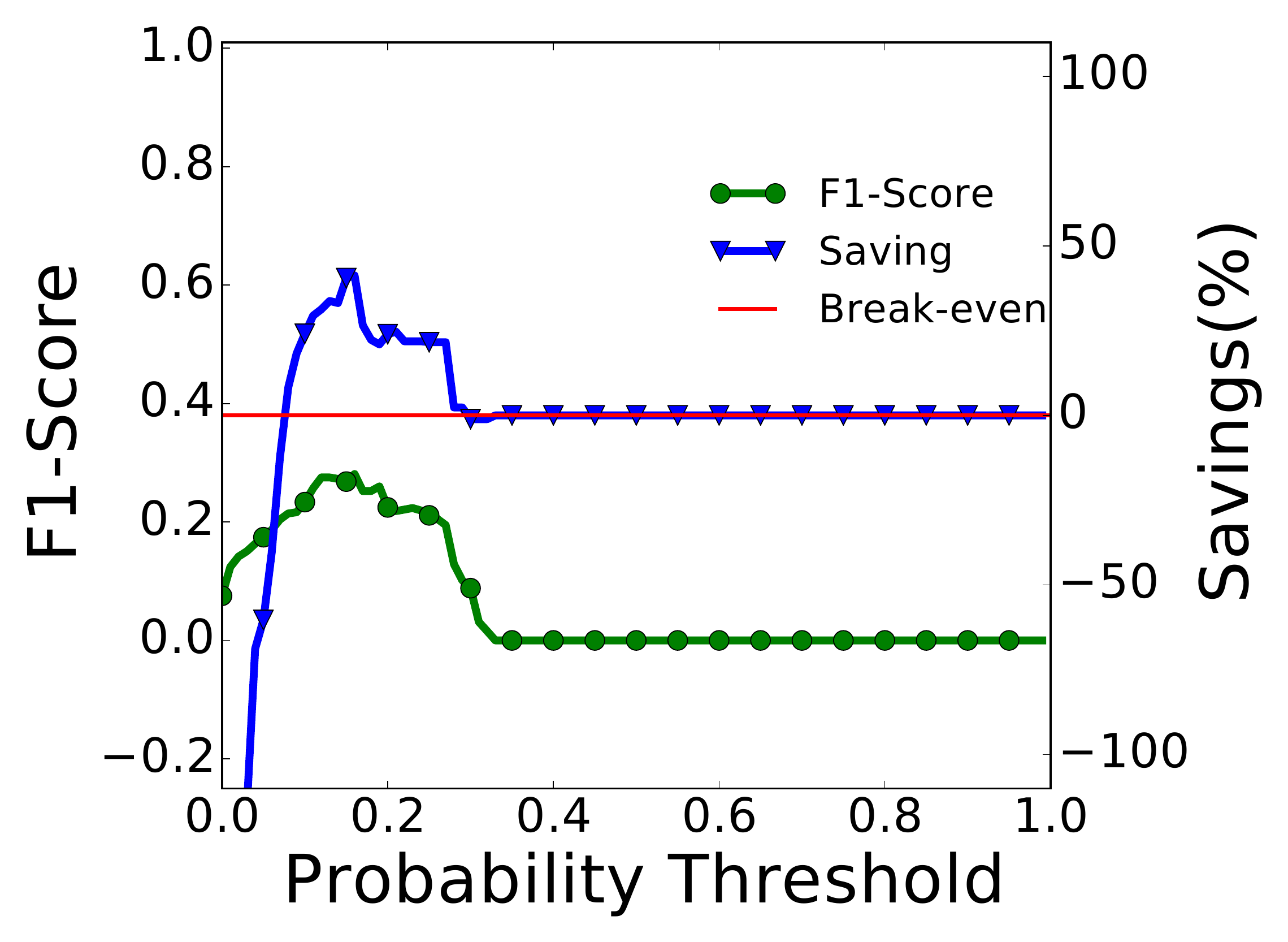}
		\caption{NLR}
		\label{fig1:F1LRSaving1}
	\end{subfigure}%
	\begin{subfigure}{.25\textwidth}
		\centering
		\includegraphics[width=1\linewidth]{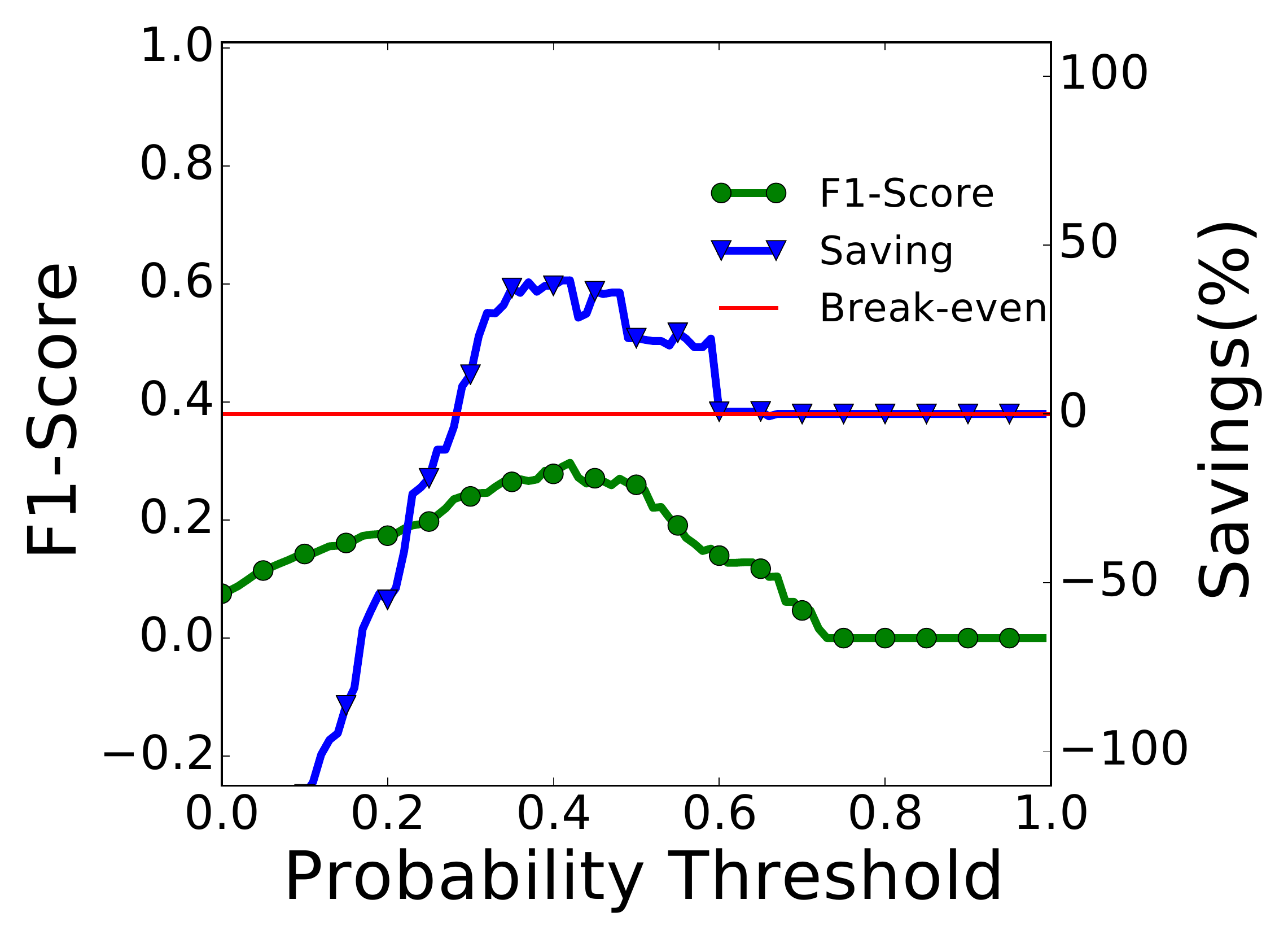}
		\caption{LR}
		\label{fig1:F1LRWSaving1}
	\end{subfigure}%
	\caption{Average savings from demand flexibility - for various F1-score. (Hourly)}
\end{figure}

\begin{figure}[!htb]
		\centering
		\includegraphics[width=.7\linewidth]{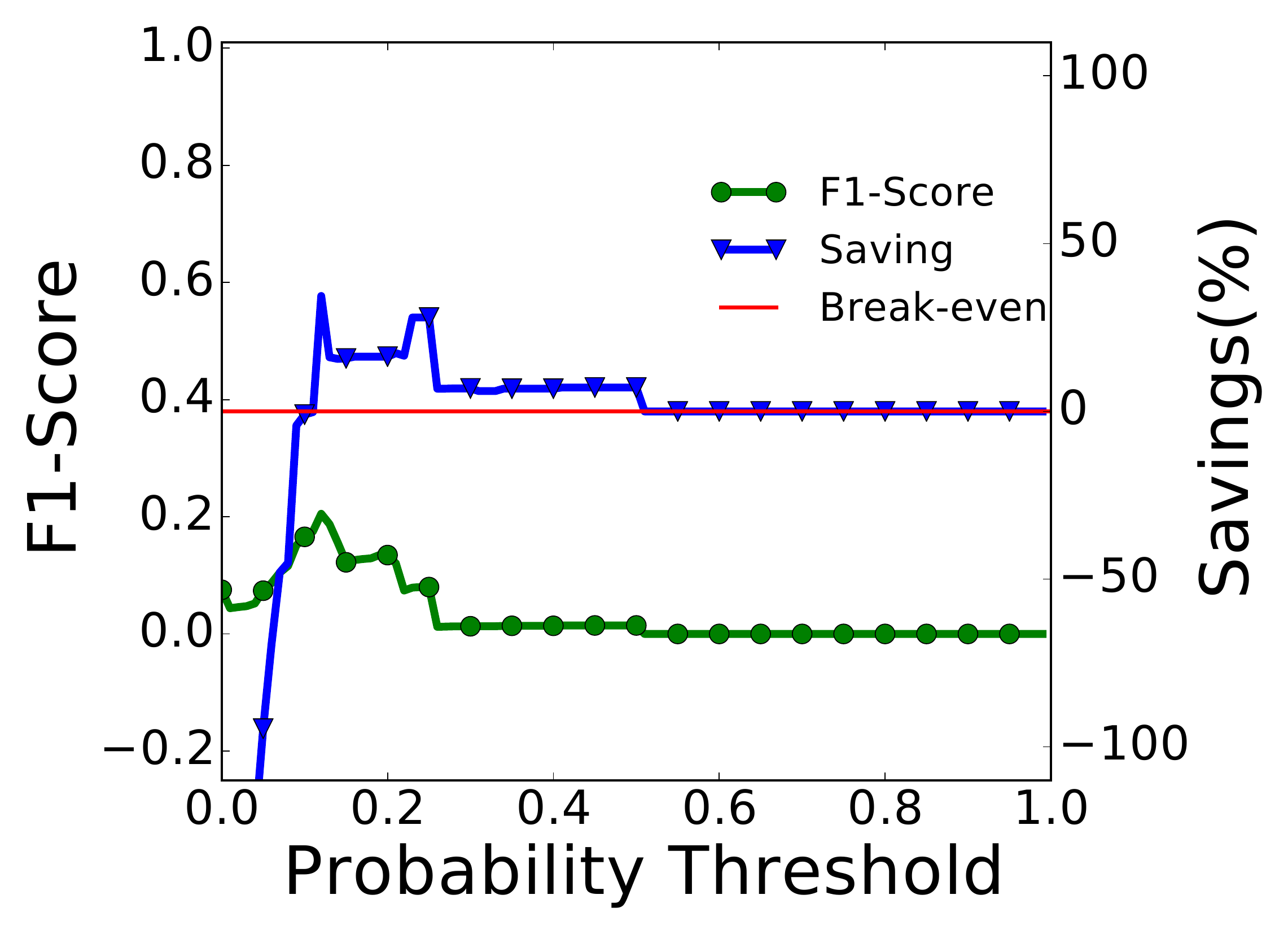}
		\caption{Average savings from demand flexibility - for various F1-score. (PM , Hourly)}
		\label{fig1:F1PMSaving1}
\end{figure}

As discussed in Section 6.3,  due to the fluctuating nature of the precision-recall curve,  selecting a probability threshold value that guarantees positive savings for a model is a difficult task. Hence, we proposed a rule of thumb of selecting the threshold value with the highest \textit{F1-score}. To further support this , we evaluate the savings relative to the \textit{F1-score} for all forecast models, shown in Figures \ref{fig1:F1LRSaving1}, \ref{fig1:F1LRWSaving1}, and \ref{fig1:F1PMSaving1}.  The figures show that for all the models, the savings follows the respective \textit{F1-score}, and the model achieves positive savings at a point of the highest \textit{F1-score}.  Thus, these results further support our arguments that the problem of setting the optimal probability threshold can be solved by selecting the threshold value with the highest \textit{F1-score}. The highest \textit{F1-score} achieved by the NLR, LR, and PM are 0.28, 0.3, and 0.21, respectively.  

\begin{figure}
	\centering
	\begin{subfigure}{.25\textwidth}
		\centering
		\includegraphics[width=1\linewidth]{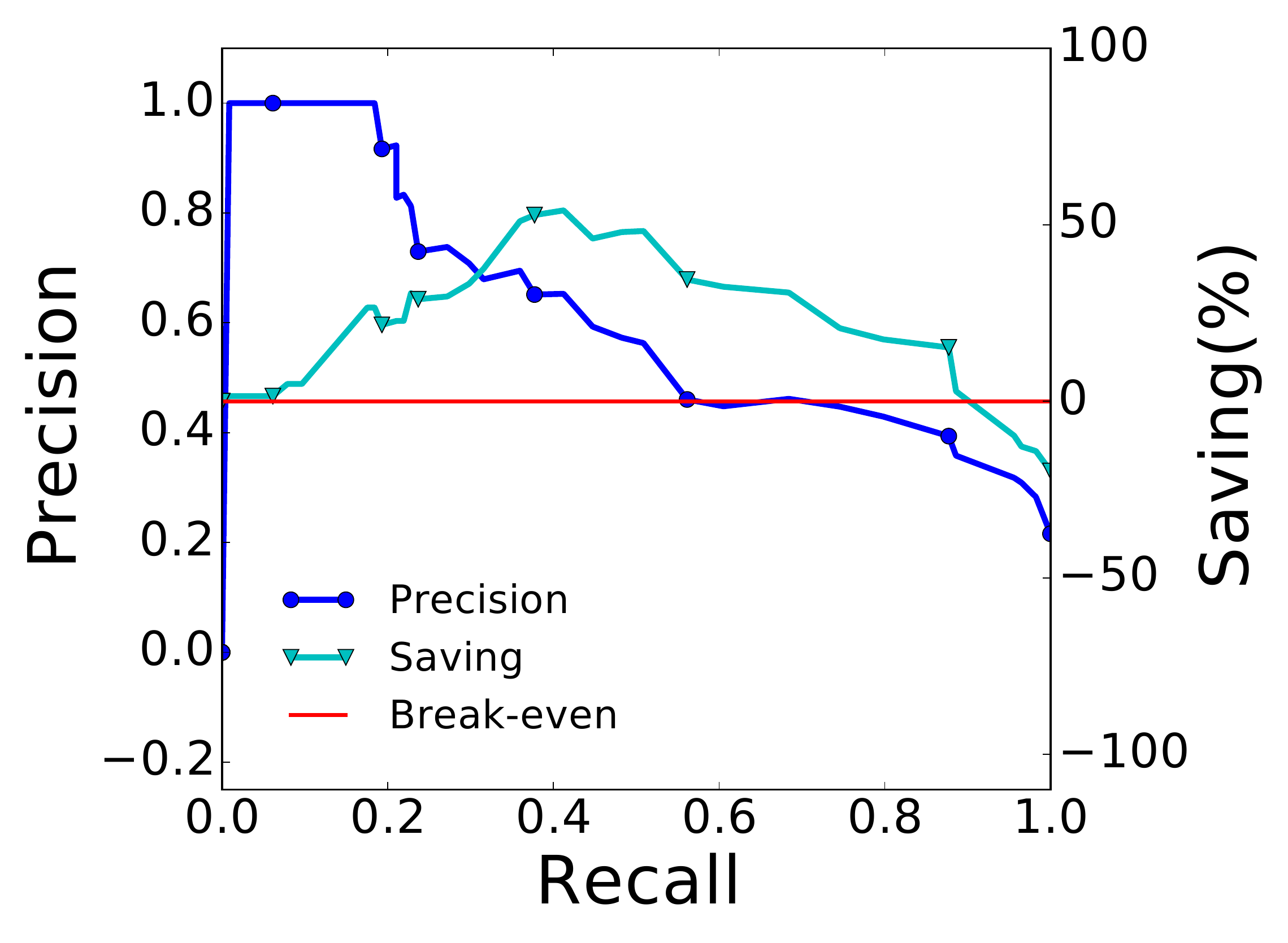}
		\caption{Group}
		\label{fig1:prLRSavingGroup1}
	\end{subfigure}%
	\begin{subfigure}{.25\textwidth}
		\centering
		\includegraphics[width=1\linewidth]{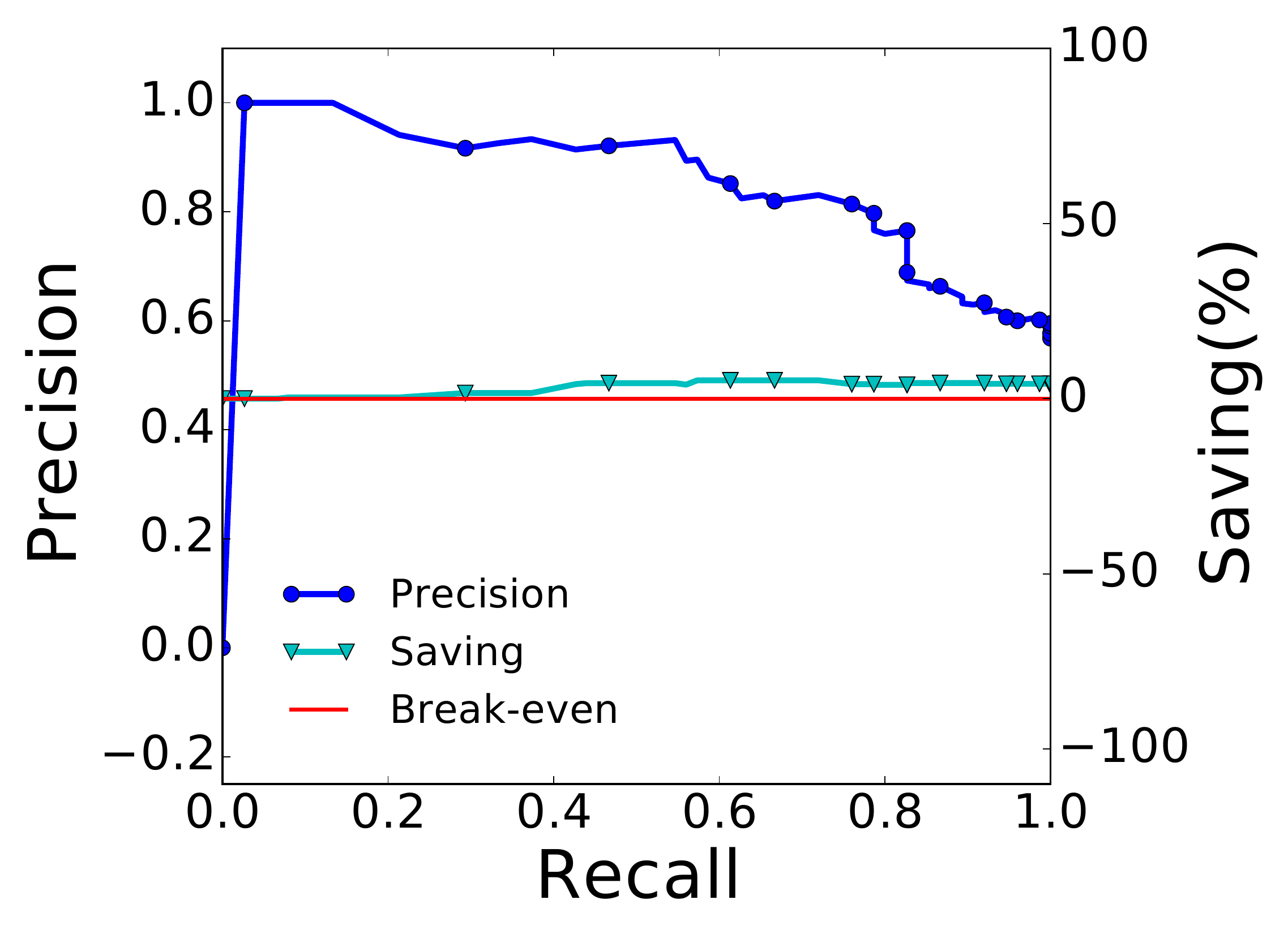}
		\caption{Daily}
		\label{fig1:prLRSavingDaily1}
	\end{subfigure}%
	
	\caption{Savings from demand flexibility of a device- relative to precision-recall (NLR). }
\end{figure} 

\begin{figure}
	\centering
	\begin{subfigure}{.25\textwidth}
		\centering
		\includegraphics[width=1\linewidth]{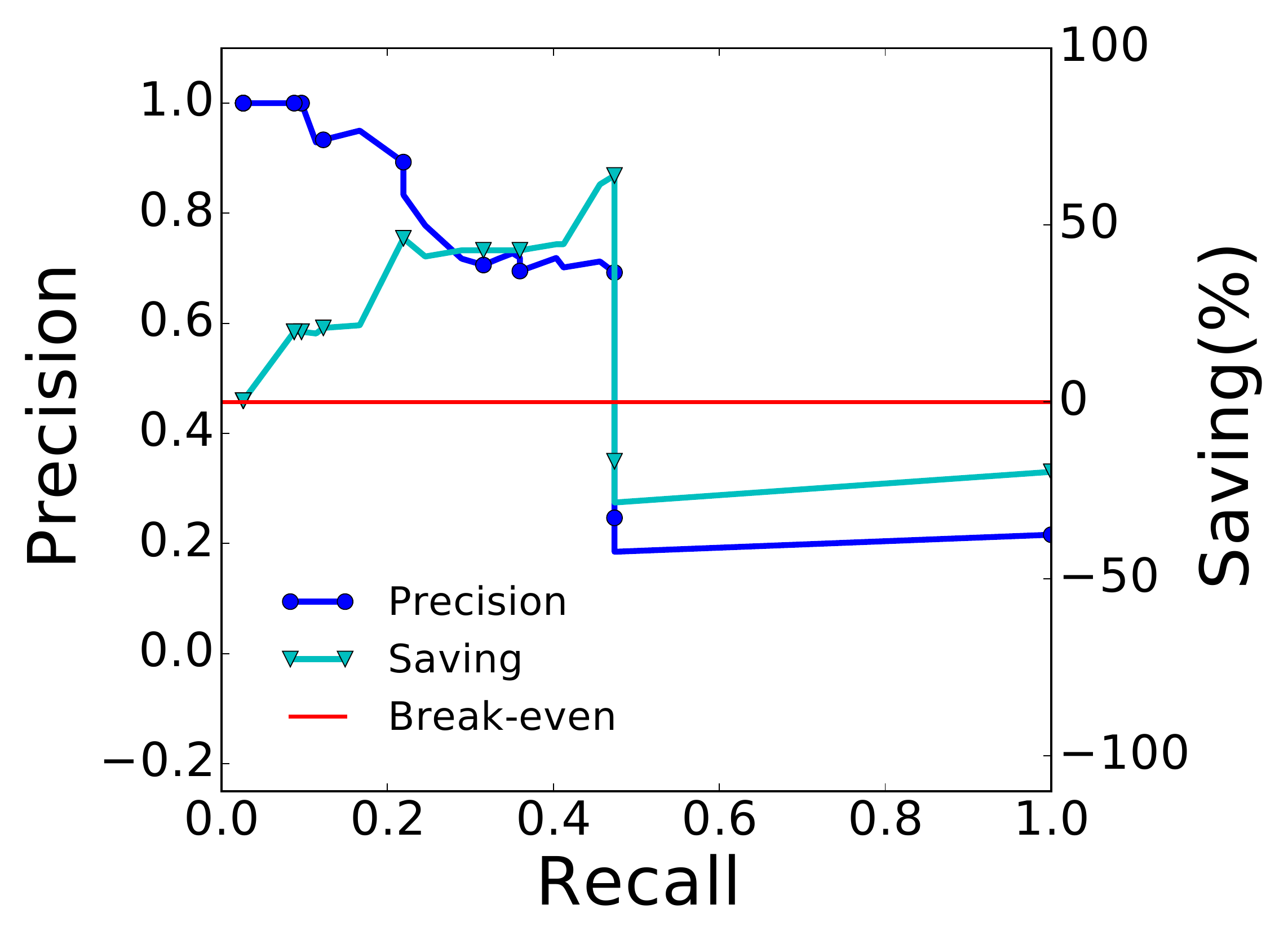}
		\caption{Group}
		\label{fig1:prPMSavingGroup1}
	\end{subfigure}%
	\begin{subfigure}{.25\textwidth}
		\centering
		\includegraphics[width=1\linewidth]{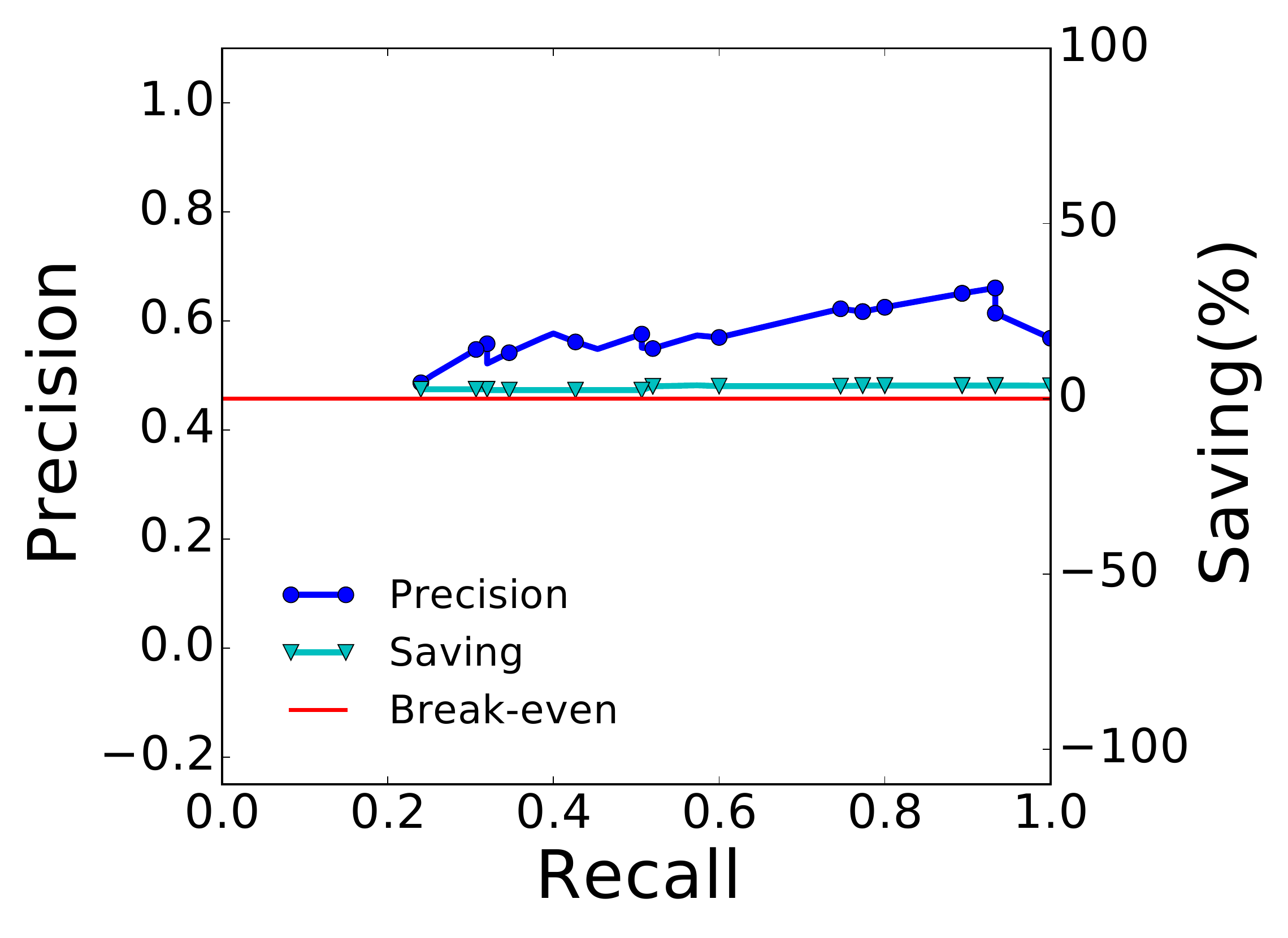}
		\caption{Daily}
		\label{fig1:prPMSavingDaily1}
	\end{subfigure}%
	\caption{Savings from demand flexibility of a device- relative to precision-recall (PM). }
\end{figure} 

To further evaluate the relationship between forecast accuracy and savings, we analyze savings from the demand forecast at the group resolution, shown in Figures \ref{fig1:prLRSavingGroup1}, \ref{fig1:prLRWSavingGroup}, and \ref{fig1:prPMSavingGroup1}. As before, for all the models the savings increases with the data aggregation. The figures show that the best \textit{F1-score} for NLR is double of hourly resolution and LR achieves 93\%  increase in savings. Because the best savings for PM lies at the point of the best \textit{F1-score}, PM surpasses the savings from LR by 10\% despite having lower prediction accuracy (AUC). For other models, best savings does not coincide with the best \textit{F1-score}. Thus, they have a comparatively lower precision and recall value at the point of the best savings. These results show that in the flexibility market, a model with a better forecast performance (AUC) does not guarantee higher savings. Further, a comparable savings from the baseline model (PM) supports our argument that, at the device-level, a simple model can compete with a complex one. Figures \ref{fig1:prLRSavingDaily1}, \ref{fig1:prLRWSavingDaily}, and \ref{fig1:prPMSavingDaily1} compares the savings at the daily resolution.  Similar to the case with LR, the savings for the other two models are also reduced due to a decrease in the number of available flexible demands to be scheduled. 

\end{document}